\documentclass[traditabstract]{aa}
\usepackage{txfonts}
\usepackage{color}
\usepackage{graphicx}
\usepackage{natbib}
\bibpunct{(}{)}{;}{a}{}{,} 

\title{Cosmic radio dipole from NVSS and WENSS}

 \author{Matthias Rubart\thanks{\tt matthiasr at physik dot
uni-bielefeld dot de},
   		Dominik J. Schwarz\thanks{\tt dschwarz at physik dot
uni-bielefeld dot de}}

\institute{Fakult\"at f\"ur Physik, Universit\"at Bielefeld, Postfach 100131, 33501 Bielefeld, Germany}

\abstract
{We use linear estimators to determine the magnitude and direction of the cosmic radio dipole from the NRAO VLA Sky Survey (NVSS) and the Westerbork Northern Sky Survey (WENSS). We show that special attention has to be given to the issues of bias due to shot noise, incomplete sky coverage and masking of the Milky Way. We compare several different estimators and show that conflicting claims in the literature can be attributed to the use of different estimators. We find that the NVSS and WENSS estimates of the cosmic radio dipole are consistent with each other and with the direction of the cosmic microwave background (CMB) dipole. We find from the NVSS a dipole amplitude of $( 1.8 \pm 0.6) \times 10^{-2}$ in direction $(\mathrm{RA}, \mathrm{dec})= (154^\circ \pm  19^\circ, -2^\circ \pm  19^\circ)$. This amplitude exceeds the one expected from the CMB by a factor of about $4$  and is inconsistent with the assumption of a pure kinetic origin of the radio dipole at $ 99.6 \%$ CL.}


\titlerunning{Radio dipole estimates}
\authorrunning{Matthias Rubart \& Dominik J.~Schwarz}

\begin{document}
\maketitle

\section{Introduction}

The assumed isotropy and homogeneity of the Universe at large scales is fundamental to modern 
cosmology. The isotropy is best seen in the cosmic microwave background (CMB) radiation and holds 
at the per cent level. The most prominent anisotropy of the CMB temperature is a dipole signal of 
$\Delta T/T \approx 10^{-3}$. It is commonly assumed that this dipole is largely caused by the motion of the 
Solar system through the Universe \citep{CMB_dipole_theory}.  This interpretation seems to be fully
consistent with the concordance model of cosmology. 

However, the observation of the microwave sky is not enough to tell the difference between a motion 
induced CMB dipole and dipole contributions form other physical phenomena, i.e.
\begin{equation}
\vec d_{\rm cmb} = \vec d_{\rm motion} + \vec d_{\rm primordial} + \vec d_{\rm ISW} + 
\vec d_{\rm foregrounds} + \vec d_{\rm noise}. 
\end{equation} 
 In our notation a dipole vector $\vec d$ modulates the isotropic sky by a factor $(1+\vec d \cdot \vec{\hat{r}})$,
with $\vec{\hat{r}}$ denoting the position on the sky.

Usually it is assumed that the primordial and the integrated Sachs-Wolfe (ISW) contribution to the 
CMB dipole are negligibly small and that foregrounds (the Milky Way) are under control. 
Within the concordance model we expect a primordial contribution of 
$d_{\rm primordial} \approx 2 \times 10^{-5}$. The ISW contribution could be as large 
as $10^{-4}$ from the gravitational potentials induced by local 100 Mpc sized structures, without being 
in conflict with the concordance model \citep{Rakic,Francis}. The noise term can be ignored 
due to excellent statistics of full sky observations. Thus the measured $\vec d_{\rm cmb}$ is 
directly used to infer the velocity of the Solar system w.r.t.~the CMB to be $v = 369 \pm 0.9 \ \rm km\ s^{-1}$  
\citep{CMB_dipole_obs}.  It is used in many cosmological studies done in the CMB rest frame, e.g. 
supernova Hubble diagrams or measurements of large scale bulk flows.

The effects of motion are not limited to the CMB, but should actually be detectable at any frequency. In order 
to test the hypothesis $\vec d_{\rm cmb} = \vec d_{\rm motion}$, it would be very interesting to measure the 
dipole of another cosmic probe, such as that obtained by radio point source catalogues. In 
this case one expects to find 
\begin{equation}
\vec d_{\rm radio} = \vec d_{\rm motion} + \vec d_{\rm structure} + \vec d_{\rm foregrounds} + \vec d_{\rm noise}. 
\end{equation}  
Besides the signal from our proper motion, we expect a signal from structure in the Universe 
and we expect a random dipole from Poisson noise. The dipole from structure is expected to dominate 
any catalogue limited to redshift $z \ll 1$. Thus we are interested in surveys with a mean redshift of 
order unity and a large enough sky coverage to be sensitive to the dipole.  This makes radio 
catalogues the preferred probe to look at. Within the concordance model, the dipole signal induced by 
the large scale structure is then a subdominant contribution, as it is for the CMB. 
If we had a large enough catalogue, we could compare $\vec d_{\rm radio}$ to $\vec d_{\rm cmb}$.  
Any statistically significant deviation would be exciting, while finding a match would put the concordance 
model on firmer grounds. 

A first attempt to measure the radio dipole was performed by \citet{Baleisis} using a combination of the 
Green Bank 1987 and the Parkes-MIT-NRAO catalogues. \citet{BW02}, \citet{Singal11} and 
\citet{Gibelyou} attempted to determine the dipole vector in the NRAO VLA Sky Survey (NVSS), 
with different conclusions. \citet{BW02} found a result that is in agreement with a purely kinetic origin 
of the cosmic radio dipole, but this was challenged by \citet{Singal11}, who finds a dipole 
amplitude four times larger than expected, but strangely enough pointing in a direction consistent with 
the CMB dipole. The analysis of  \citet{Gibelyou} finds both a different direction and an amplitude six
times as large as the expected one. While \citet{BW02} used a quadratic estimator, \citet{Singal11} and 
\citet{Gibelyou} used different linear estimators to find the dipole direction.  

The purpose of this work is to discuss the use of linear estimators of the cosmic radio dipole and apply
several versions of them on the NVSS \citep{NVSS} and the Westerbork Northern Sky Survey 
(WENSS) \citep{WENSS}. 
We resolve the conflicts in the literature and extend the analysis to other linear estimators. 

The NVSS survey covers about $10.3$ sr of the sky and contains about $2 \times 10^5$ 
sources per steradian.  For this survey the Very Large Array (VLA) in New Mexico (USA) has been 
used measuring at a frequency of $1.4$ GHz. The survey includes over $80$ per cent of the 
sky, missing only areas with declination $\delta < -40^\circ$. The lower flux limit lies at 
$2.5$ mJy for the $5 \sigma$ detection of point sources. The NVSS was conducted by means of 
two different configurations of the VLA above and below $\delta = -10^\circ$. 

The Westerbork Synthesis Radio Telescope in the Netherlands 
was operated at a frequency of $325$ MHz
to record the WENSS survey covering about $2.9$ sr of the nothern sky and containing about 
$2.3 \times 10^5$ sources in total. This survey is made up of a main 
catalog for  $\delta \in (28^\circ, 76^\circ)$ and a polar catalog  for $\delta > 72^\circ$. The 
$5 \sigma$ detection limit for this survey is $18$ mJy. 

In order to analyze these surveys, we focus on linear estimators in this work. We do so for two 
reasons. Firstly, recent controversial results used linear estimators for the dipole direction 
\citep{Singal11,Gibelyou} and in one work also for the dipole amplitudes \citep{Singal11}. Secondly, 
linear estimators are conceptually simpler. However, it is not expected that they are optimal (unbiased 
and minimal variance). The linear estimators used in our analysis are asymptotically unbiased
and their variance can be easily understood by analytic calculations and by Monte Carlo simulations.    

The paper is organized as follows: First we discuss the expected kinetic radio dipole. 
In section \ref{Previous results} we outline previous estimates of the radio dipole. 
Linear estimators for full sky surveys are investigated in section \ref{AmplitudenBias}, 
followed by a detailed analysis of the effects of incomplete sky coverage and masking in 
the next section. In section \ref{xProblem} we discuss the expected dipole amplitude from a 
flux based estimator. Our estimate of the radio dipole can be found in section \ref{Ourresults} and 
is followed by a comparison with previous results. We conclude in section \ref{Conclusion}.

\section{Kinetic Radio Dipole}

\subsection{Doppler shift and aberration} 

\citet{1984} predicted the kinetic contribution to the cosmic radio dipole for an isotropic and 
homogeneous cosmology.  At redshift of order unity and beyond, we expect this kinetic contribution 
to be the dominant one. 

The spectrum of   a radio source is assumed to be described by a power law, 
\begin{equation}
\label{fluxpowerlaw}
S(f) \propto f^{-\alpha}, 
\end{equation}
where $S$ denotes the flux and $f$ the frequency. Each radio source can be described by an individual
 spectral index $\alpha$. For simplicity we assumed a mean value of $\alpha$ for all radio sources in the catalogue. 

The number of observed radio sources per steradian depends on the lower flux limit and can 
be approximated by a power law
\begin{equation}
\label{numbercountpowerlaw}
 \frac{dN}{d\Omega}(>S) \propto S^{-x} .
\end{equation}
The value of $x$ can be different for each survey.  Typically $x$ is assumed to be about one. 

Two effects have to be taken into account. The emitted radio frequency $f_\mathrm{rest}$ is 
observed at the Doppler shifted frequency $f_\mathrm{obs}$. The magnitude of this change depends on 
the angle $\theta$ between the direction to the source and the direction of our motion, with velocity $v$. Observed 
and rest frame frequencies are related by
\begin{equation}
 f_\mathrm{obs} = f_\mathrm{rest} \delta(v,\theta) ,
\end{equation}
where $\delta$ is given by
\begin{equation}
\label{deltaexact}
 \delta(v,\theta) = \frac{1+\frac{v}{c} \cos(\theta)}{\sqrt{1-(\frac{v}{c})^2}} .
\end{equation}
Thus the observed flux changes due to our motion, since it depends on the frequency
\begin{equation}
\label{fluxchange}
S_{\mathrm{obs}}(f_{\mathrm{obs}}) \propto  \delta f_{\mathrm{rest}}^{-\alpha} \propto 
\delta^{1+\alpha} f_{\mathrm{obs}}^{-\alpha} \propto S_{\mathrm{rest}} (f_{\rm obs}) \delta^{1+\alpha}.
\end{equation}
The first factor of $\delta$ is due to the fact that the energy of an observed photon is enhanced due to 
the Doppler effect. 

Thus, the Doppler effect will change the number of observed sources above a given 
flux limit like
\begin{equation}
 \left(\frac{dN}{d\Omega}\right)_\mathrm{obs}=\left(\frac{dN}{d\Omega}\right)_\mathrm{rest} \delta^{x(1+\alpha)}.
\end{equation}

Since the velocity of light is finite, aberration will also modify the number counts. 
The position of each source is changed towards the direction of motion. The new angle 
$\theta^\prime$ (observed from Earth) between the position of the source and the direction of 
motion is given by 
\begin{equation}
\label{thetaprime}
 \tan{\theta^\prime} = \frac{\sin{\theta}\sqrt{1-\frac{v^2}{c^2}}}{\frac{v}{c}+\cos{\theta}} .
\end{equation}
Therefore, at first order in $v/c$, $d\Omega$ transforms like
\begin{equation}
 d \Omega^\prime = d \Omega (1-2\frac{v}{c} \cos{\theta}) + O\left( \left(\frac{v}{c}\right)^2\right)  .
\end{equation}
This can be combined with the Doppler effect to give the observed number density. After approximating $\delta(v,\theta)$ to first order in $\frac{v}{c}$, the result becomes
\begin{equation}
\label{numbercountapprox}
 \frac{dN}{d\Omega}_{\mathrm{obs}} = \left(\frac{dN}{d\Omega}\right)_{\mathrm{rest}}  \left[1+[2+x(1+\alpha)] \left(\frac{v}{c}\right) \cos(\theta)\right] .
\end{equation}
The amplitude of the kinetic radio dipole is then given by
\begin{equation}
\label{amplitudedefinition}
d = [2+x(1+\alpha) ] \left(\frac{v}{c}\right). 
\end{equation}
The kinetic radio dipole points towards the direction of our peculiar motion, which in an isotropic 
and homogeneous Universe must also agree with the direction defined by the CMB dipole.

\subsection{Expected kinetic radio dipole}

The measured CMB dipole is $\Delta T = 3.355 \pm 0.008$ mK  in the direction 
(l,b) = ($263.99^\circ \pm 0.14^\circ, 48.26^\circ \pm 0.03^\circ$) \citep{CMB_dipole_obs}.
In equatorial coordinates (epoch J2000) its direction reads (RA,dec) $=  (168^\circ, -7^\circ)$.
Compared to the CMB temperature of $T_0 = 2.725 \pm 0.001$ K \citep{Fixsen}. 
this corresponds to a relative fluctuation of $\Delta T/T = (1.231\pm 0.003) \times 10^{-3}$ and thus 
the velocity of the Solar system has been inferred from the CMB dipole 
to be $v = 369.0 \pm 0.9\ \rm km\ s^{-1} $ \citep{CMB_dipole_obs}. 

\begin{figure}[!ht]
\begin{center}
 \includegraphics[width=6cm,angle=270]{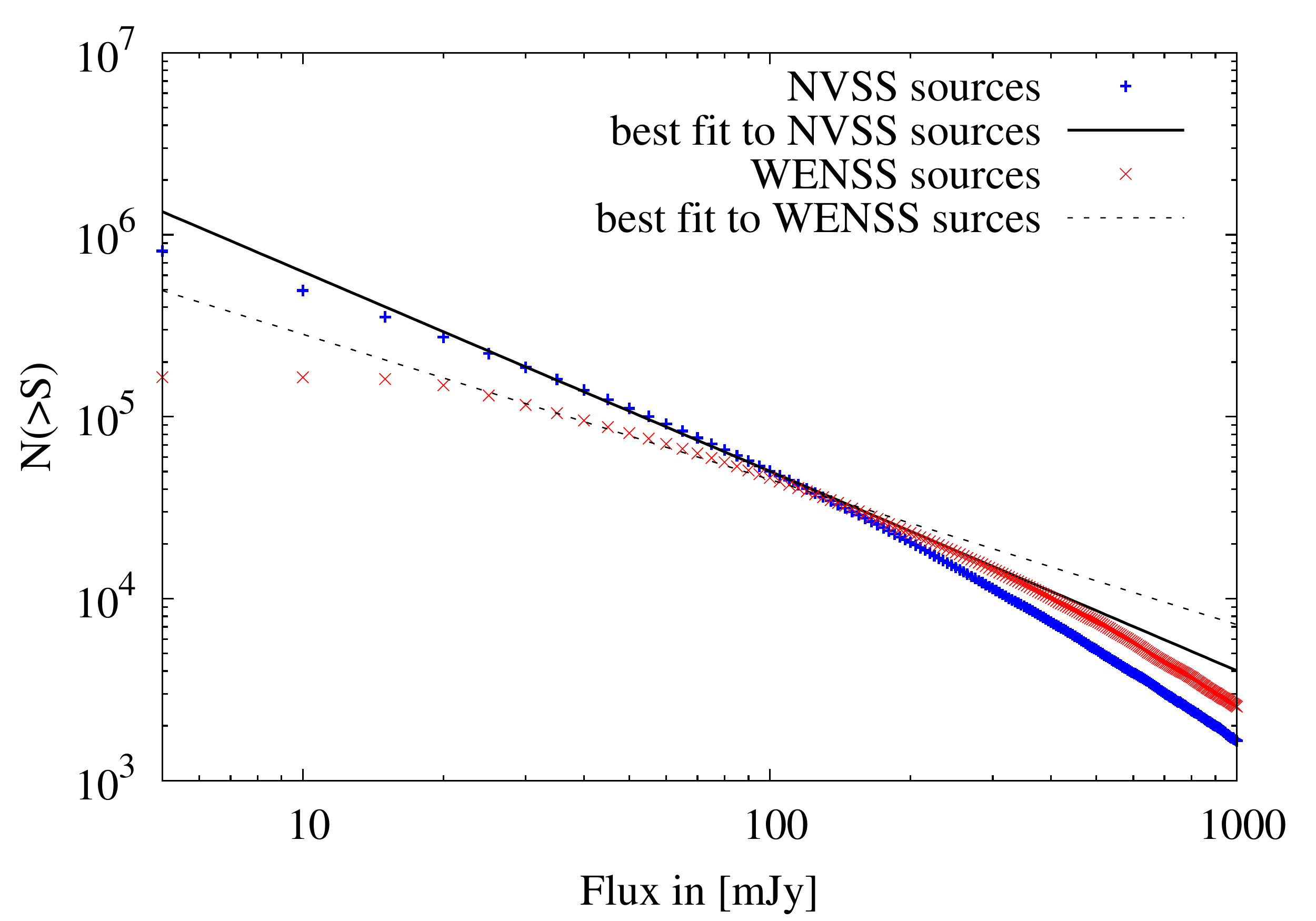}
\caption{Number counts of the NVSS and WENSS surveys. A function $f(S) \propto S^{-x}$ is fitted to both data sets in the range of $25\ \mathrm{mJy} < S < 200\ \mathrm{mJy}$. Resulting values of x are $1.10 \pm 0.02$ for the NVSS survey and $0.80 \pm 0.02$ for the WENSS survey.}\label{bothnbiggers}
\end{center}
\end{figure}

In order to find the expected amplitude of the kinetic radio dipole, we also need estimates for 
$x$ and $\alpha$. The typically assumed values are $x = 1$ and $\alpha = 0.75$, which 
gives together with $v = 370 \  \rm km\ s^{-1}$ a radio dipole amplitude of $d = 0.46 \times 10^{-2}$. 
However, we can improve on that as $x$ can be measured with help of the radio survey.  Therefore we need to
plot $N(>S)$ against $S$ like in figure \ref{bothnbiggers}. 

For the purpose of this work we find $x_{\rm NVSS} = 1.10 \pm 0.02$ and 
$x_{\rm WENSS} = 0.80 \pm 0.02$.  The mean spectral index cannot be inferred from the catalogues, as they 
provide data at a single frequency band only. We thus stick to $\alpha = 0.75$, but include in the 
dipole error an uncertainty of $\Delta \alpha = 0.25$ \citep{alpharef}. This results in the expectations:
\begin{eqnarray}
d_{\rm NVSS}^{\rm exp}      &=& (0.48 \pm 0.04) \times 10^{-2}, \\
d_{\rm WENSS}^{\rm exp}   &=& (0.42 \pm 0.03) \times 10^{-2}.
\end{eqnarray}
The error is dominated by the uncertainty in the spectral index. 
 
\section{Previous results}
\label{Previous results}

The first measurement of the radio dipole using the NVSS catalogue was performed by \citet{BW02}. 
In order to remove corruption by local structure, all sources within $15^\circ$ vicinity of 
the galactic disk have been removed. Additionally the clustering dipole contribution was reduced by 
ignoring sources within $30^{\prime \prime}$ of nearby known galaxies. The spherical harmonic 
coefficients $a_{lm}^{\mathrm{obs}}$ from the remaining NVSS catalogue 
have been determined up to $l=3$. A model for a dipole distribution with an isotropic background has 
been constructed ($a_{00}$ and $a_{10}$). Due to masking, this dipole distribution also influences 
higher multipoles. After applying the same mask as for the NVSS catalogue, one finds 
$a_{lm}^{\mathrm{model}}$ up to $l=3$. A quadratic estimator (chi square) was used to compare the 
model with the observed coefficients. 

The resulting best-fit dipoles can be seen in table \ref{BWResult}. The results of \citet{BW02} indicate a 
higher radio dipole than expected, however without statistical significance. 

\begin{table}[h!]
\begin{center}
\begin{tabular}{cccccc}
\hline
Flux & $N$ & RA & dec & $d$ & $\chi_{red}^2$  \\
(mJy) &  &  ($^o$) &  ($^o$) & $(10^{-2})$&  \\
\hline
$> 40$ & 125,603 & 149 $\pm$ 49 & -45 $\pm$ 38 & 0.7 $\pm$ 0.5  & 1.02 \\
$> 35$ & 143,524 & 161 $\pm$ 44 & -27 $\pm$ 39 & 0.9 $\pm$ 0.4 & 0.74 \\
$> 30$ & 166,694  & 156 $\pm$ 32 & 2 $\pm$ 33 & 1.1 $\pm$ 0.4& 1.01 \\
$> 25$ & 197,998 & 158 $\pm$ 30 & -4 $\pm$ 34 & 1.1 $\pm$ 0.3 & 1.01 \\
$> 20$ & 242,710 & 153 $\pm$ 27 & -3 $\pm$ 29 & 1.1 $\pm$ 0.3 & 1.32 \\
$> 15$ & 311,037  & 148 $\pm$ 29 & 31 $\pm$ 31 & 0.8 $\pm$ 0.3& 1.81 \\
$> 10$ & 431,990  & 132 $\pm$ 29 & 65 $\pm$ 19 & 0.5 $\pm$ 0.2& 4.96 \\
\hline
\end{tabular}
\caption{Best-fit dipole parameters from \citet{BW02}. Coordinate system and amplitude definition 
are adjusted for comparison with Singal's results (see tables \ref{SingalresultsN} and \ref{Singalresults}).
$N$ denotes the number of point sources with flux above the indicated limit. \label{BWResult}}
\end{center}
\end{table}

\citet{Singal11} used a linear estimator, originally proposed by \citet{Crawford},
\begin{equation}
\label{Nsum}
 \vec R_{\mathrm{3D}} = \sum \vec{\hat r}_i ,
\end{equation}
and a variation of it, which we discuss below. 
For a large number of sources the isotropic background will clear away. The remaining vector 
$\vec R_{\mathrm{3D}}$ will point towards the main anisotropy in the distribution of number density over the 
sky. To get the correct dipole amplitude $d$ one has to normalize this estimator depending on the number of 
sources. In Singal's analysis sources within $10^\circ$ of the galactic plane have been 
removed. In order to avoid a directional bias (see the more detailed discussion below) he reestablished a 
north-south symmetry of the NVSS by cutting all sources with $\rm dec > 40^\circ$. The results of 
\citet{Singal11} are shown in table \ref{SingalresultsN}.
The errors of the directional measurements are quite small here. This is an effect of an 
unexpectedly large amplitude, which simplifies the measurement. While the direction agrees with the one 
found by \citet{BW02}, the dipole amplitude seems to be a factor of about four higher than expected 
from the CMB dipole and twice as big as found by \citet{BW02}. 

Masking the supergalactic plane in order to reduce the contribution of local structure did not resolve the 
discrepancy.
Since unknown clustering further away from the super galactic plane could also have contributed to the 
measurement, a second test was performed. A clustering contribution to the dipole would not give a 
signal proportional to $\cos \theta$. On the other hand, the difference in number counts of areas that 
are opposite to each other should decrease with $\cos \theta$ (where $\theta$ is the angle between 
an area and the measured dipole direction), if the measured dipole is due to our velocity. Singal was 
able to fit such a behaviour to the data. Therefore he concludes that the radio dipole amplitude is not 
due to local clustering . 

\begin{table}[h!]
\begin{tabular}{ccccc}
\hline
 Flux & N  &  RA & dec & $d$   \\
 (mJy) &  & ($^{\circ}$)& ($^{\circ}$)& ($10^{-2}$) \\ \hline
$ \geq 50$ & $91,597$  & $171 \pm 14$ & $-18 \pm 14$ & $2.1\pm 0.5$ \\
$ \geq 40$ & $115,837$  & $158 \pm 12$ & $-19 \pm 12$ & $1.8\pm 0.4$  \\
$ \geq 35$ & $132,930$  & $157 \pm 11$ & $-12 \pm 11$ & $1.9\pm 0.4$ \\
$ \geq 30$ & $154,996$  & $156 \pm 11$ & $-02 \pm 10$ & $2.0\pm 0.4$ \\
$ \geq 25$ & $185,474$  & $158 \pm 10$ & $-02 \pm 10$ & $1.8\pm 0.4$  \\
$ \geq 20$ & $229,365$  & $153 \pm 10$ & $+02 \pm 10$ & $1.8\pm 0.3$ \\
$ \geq 15$ & $298,048$  & $149 \pm 09$ & $+15 \pm 09$ & $1.6\pm 0.3$ \\
\hline
\end{tabular}
\caption{Dipole direction and amplitude from the number count estimator (\ref{Nsum}) from 
\citet{Singal11}. \label{SingalresultsN}} 
\end{table}

\cite{Singal11} also used a linear estimator for the distribution of flux over the sky. This estimator is 
similar to the number density estimator (\ref{Nsum}), but weights each radio source by its flux $S_i$,
\begin{equation}
\label{Ssum}
 \vec R_{\mathrm{flux}} = \sum S_i \vec{\hat r}_i  .
\end{equation}
Like $\vec R_{\mathrm{3D}}$, this estimator finds the main anisotropy and the amplitude 
needs to be normalized. The brightest sources ($S>1000$ mJy) are removed, because they 
would dominate $\vec R_{\mathrm{flux}}$ otherwise. Results of this estimator are shown 
in table \ref{Singalresults}. The estimated directions are in agreement with the results 
of \citet{BW02} and the number count estimator results of \citet{Singal11}. However, the 
normalized dipole amplitudes $d$ are even higher than those of the number count 
estimator $\vec R_{\mathrm{3D}}$. In section \ref{xProblem} we resolve this conflict.

\begin{table}[ht!]
\begin{tabular}{ccccc}
\hline
 Flux & N    &  RA & dec & $d$\\
 (mJy) & & ($^{\circ}$)& ($^{\circ}$)  & ($10^{-2}$)\\ \hline
$ 1000 > S \geq 50$ & $90,360$  & $163 \pm 12$ & $-11 \pm 11$ & $2.3\pm 0.7$  \\
$ 1000 > S \geq 40$ & $114,600$  & $159 \pm 12$ & $-11 \pm 11$ & $2.2\pm 0.6$ \\
$ 1000 > S \geq 35$ & $131,691$ & $159 \pm 11$ & $-10 \pm 10$  & $2.2\pm 0.6$  \\
$ 1000 > S \geq 30$ & $153,759$  & $159 \pm 11$ & $-07 \pm 10$  & $2.2\pm 0.6$\\
$ 1000 > S \geq 25$ & $184,237$  & $159 \pm 10$ & $-07 \pm 09$ & $2.2\pm 0.6$ \\
$ 1000 > S \geq 20$ & $228,128$  & $158 \pm 10$ & $-06 \pm 09$ & $2.1\pm 0.5$ \\
$ 1000 > S \geq 15$ & $296,811$ & $157 \pm 09$ & $-03 \pm 08$  & $2.0\pm 0.5$  \\
\hline
\end{tabular}
\caption{Dipole direction and amplitude from the flux weighted number count estimator (\ref{Ssum}) 
from \citet{Singal11}.\label{Singalresults}}
\end{table}

Most recently, \citet{Gibelyou} measured a dipole amplitude $(d=2.7 \pm 0.5) \times 10^{-2}$ 
towards (RA,dec) $=(117 \pm 20^\circ,6 \pm 14^\circ)$ from the NVSS. 
This direction is inconsistent with the studies mentioned 
above and the dipole amplitude is a factor of five larger than expected.
The authors used separate estimators for the direction and the amplitude. 
Their direction estimate is based on a linear estimator, originally proposed by \citet{Hirata},
\begin{equation}
\label{Gibelyouesti}
 \vec R_{\mathrm{3DM}}=\sum_i^{N_D} \hat{r}_i  -  \frac{N_D}{N_R} \sum_j^{N_R} \hat{r}_j .
\end{equation}

This three-dimensional estimator (3DM) is intended to be unbiased for arbitrary survey geometries 
and arbitrary masking. 
The idea is to achieve that with help of the second sum, which goes over $N_R$ randomly distributed 
points, subject to the same masking. 
Therefore, the authors include all sources of the NVSS survey, except for those within $10^\circ$ 
of the galactic plane. Below we show that this estimator has a direction bias, which depends on the 
real dipole anisotropy.

We summarize, there is no agreement on the amplitude and direction of the cosmic radio dipole so far. 

\section{Linear estimators for a full sky}
\label{AmplitudenBias}

Let us first show that the estimator (\ref{Nsum}) provides an unbiased estimate of the dipole
direction. 

Starting from the distribution of the number of radio sources per solid angle (\ref{numbercountapprox}), 
as seen by a moving observer in an otherwise isotropic Universe,  
the probability to find a given radio source within a  solid angle ${\rm d}\Omega$ of position $\vec{\hat r}$ is given by
\begin{equation}
 \label{density2}
p(\vec{\hat{r}}) {\rm d} \Omega = \frac{1}{4\pi} (1 + \vec{\hat r} \cdot \vec{d}) {\rm d}\Omega,
\end{equation}
where $\vec{d}$ denotes the dipole vector.

In order to study the bias of an estimator we calculate its expectation value with respect to an ensemble 
average. We do so below by means of Monte Carlo studies. For analytic considerations, for large $N$ we 
replace the ensemble average by a spatial average, i.e.
\begin{equation}
\langle 1 \rangle = \int \prod_{i=1}^N {\rm d} \Omega_i p(\vec{\hat r}_i) 1 = 1,
\label{measure}
\end{equation}
thus we assume ergodicity. Note that the average is a linear operator. 

Now the expectation value of Crawford's estimator can be evaluated for large $N$,
\begin{equation}
 \langle \vec{R}_{\mathrm{3D}} \rangle = \langle \sum_{i=1}^N \hat{r}_i \rangle  = 
 \sum_{i=1}^N \langle \hat{r}_i \rangle =
 \frac{N}{4 \pi}\int \mathrm{d} \Omega\,  (1 +  \hat{r} \cdot \vec{d})\, \hat{r} .
\end{equation}
This calculation holds for independent, identically distributed positions $\hat{r}_i$, thus without 
clustering effects. Only the second term survives the integration and thus the expected dipole estimator is 
\begin{equation}
 \label{expectCrawford}
\langle \vec{R}_{\mathrm{3D}} \rangle =\frac{1}{3} N \vec{ d}.
\end{equation}
Naively, one could now estimate the dipole signal by $\vec d_{\mathrm{3D}} \equiv \frac{3}{N}  
\vec{R}_\mathrm{3D} $. 

We conclude that $\vec d_{\rm {3D}}$ provides us with an unbiased estimate of the dipole direction 
$\vec{\hat d}$ for a  full sky sample. However the estimated dipole amplitude $|\vec d_{\rm{3D}} |$ is biased.

To understand the origin of this bias let us first  consider
\begin{equation}
\label{dCsquare}
  <\vec d^2_{\rm{3D}}> =  \left(1 - \frac{1}{N} \right)  d^2  + \frac 9 N > d^2.
\end{equation} 
The inequality holds for large $N$ and $d < 3$ (in case of large dipole amplitudes [$d= {\cal O}(1)$] 
our ansatz (\ref{measure}) should also take many-point correlations into account).
Thus $\vec d^2_{\rm{3D}}$ is definitely biased towards higher amplitudes. 
However, in order to prove that $|\vec d_{\rm{3D}} |$ is biased, we would need to calculate 
$\left<|\vec d _{\rm{3D}}|\right>$. We do this by means of the random walk/flight method.
 
\subsection{Random flight}

Adding up vectors for each point of a survey corresponds to a random walk with unit step size. To be more 
precise this is a random flight, since the problem is three dimensional. Even for a vanishing dipole, 
such a random flight is unlikely to return to the origin after $N$ steps. This describes the noise of 
any realisation of an isotropic distribution of $N$ sources.

Following \citet{Crawford}, we determine the distance $r$ from the origin after $N$ steps 
from the probability density of a random flight process
\begin{equation}
\label{Result3DFlight}
 \check{P}_N(r) \mathrm{d}r= \left[ \frac{54}{\pi N^3}\right]^{1/2} r^2 \exp\left( -\frac{3 r^2}{2N} \right)  
 \mathrm{d}r.
\end{equation}

The probability of measuring a dipole signal of an amplitude bigger than $R$ in a random flight is
\begin{equation}
 P_N(R>R_{pCL})= \int_{R_{pCL}}^\infty \mathrm{d} r \check{P}_N(r)=1 - pCL  .
\end{equation}
A confidence level $pCL$ can be choosen, leading to errorbars for a measured dipole vector 
$R_\mathrm{3D} \pm R_{pCL}$. To estimate the directional uncertainties of this method, 
\cite{Crawford} made the following argument: at a given confidence level the random flight corresponds 
to a step of length up to $R_{pCL}$. Adding $R_{pCL}$ perpendicular to the measured dipole 
$\vec{R}_\mathrm{3D} $ allows us to estimate the maximal offset in direction. 
Using trigonometry, one can relate $R_{pCL}$ to the directional uncertainties expressed as the
angle
\begin{equation}
 \delta \theta_{pCL} = \arcsin{\frac{R_{pCL}}{R_\mathrm{3D}}} .
\end{equation}

The expected magnitude of the random dipole contribution is estimated from (\ref{Result3DFlight}) as  
\begin{equation}
\label{estimateA}
 \langle R_{\mathrm{3D}}^{\mathrm{random}} \rangle= \int_0^\infty r \left[ \frac{54}{\pi N^3}\right]^{1/2} r^2 \exp\left( -\frac{3 r^2}{2N} \right) \mathrm{d}r  \approx 0.92 \sqrt{N}  .
\end{equation}

Since the random dipole has no distinguished direction, there is no direction bias of the linear estimator
for a full sky map.

Even for vanishing $d$, this gives rise to a non-vanishing estimate of the dipole amplitude 
$d_\mathrm{3D}^\mathrm{random} = 2.8 N^{-1/2}$ and is thus the origin of the amplitude bias. 

\begin{figure}[t]
\begin{center}
 \includegraphics[width=6cm,angle=270]{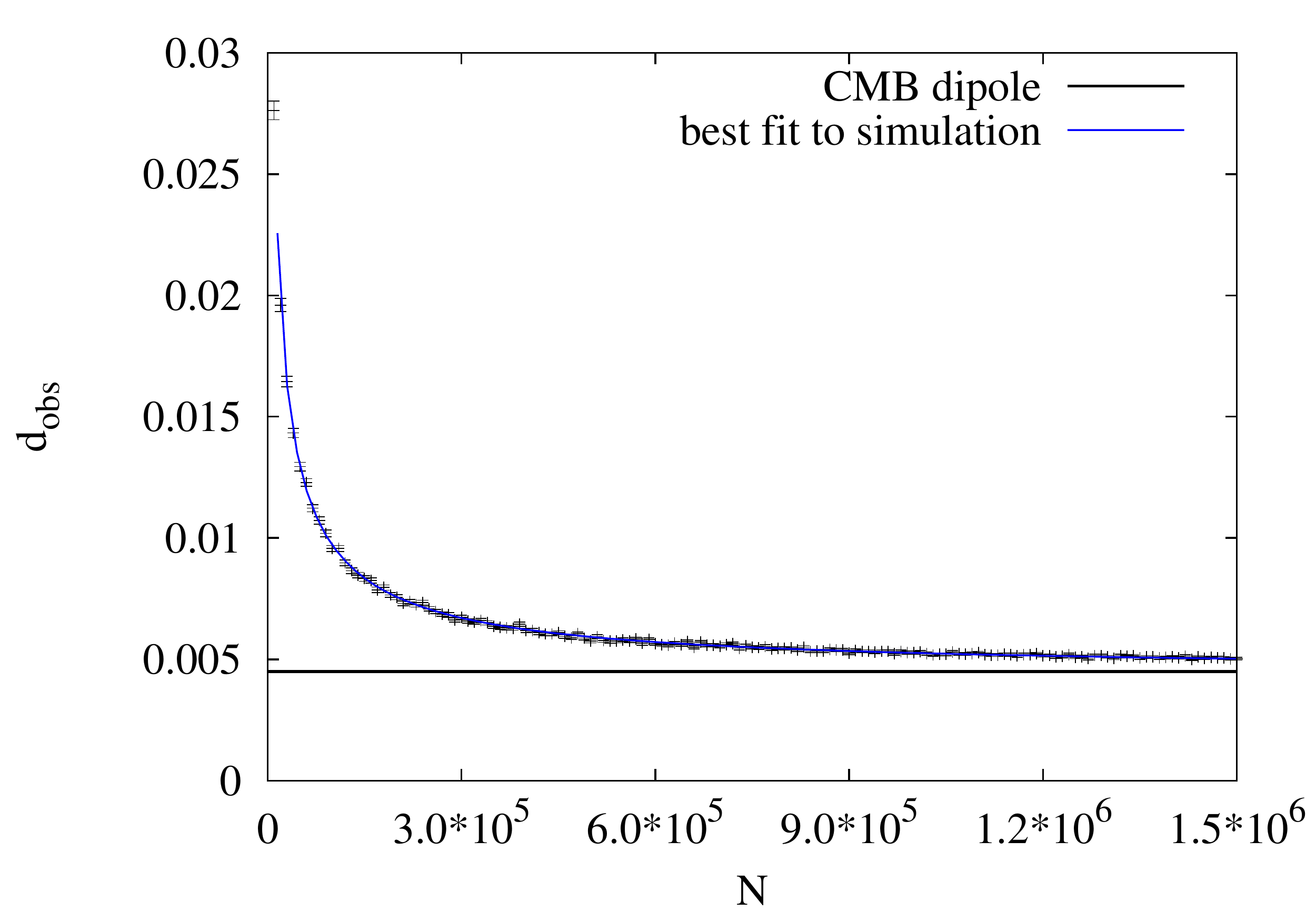}
\caption{Ampliude bias of the full sky estimator $d_{\rm{3D}}$. 
Data represent mean and empirical variance of 1000 simulations 
for each $N$. A function $d_{\rm{obs}}(N) = \sqrt{D^2 + 9 A^2/N}$ 
is fitted to the simulated data, with best-fit values 
$A= 0.908 \pm 0.002, D= (0.451 \pm 0.001) \times 10^{-2} $. The expected dipole amplitude 
($d = 0.0046$) is indicated by the horizontal line.\label{3DGraph}}
\end{center}
\end{figure}

Motivated by (\ref{dCsquare}) we make the following ansatz for the dipole amplitude and its bias: 
\begin{equation}
\label{ansatzd}
 \langle d_{\mathrm{3D}}\rangle = \sqrt{d^2 + \frac{9}{N^2} \langle R_\mathrm{3D}^\mathrm{random} \rangle^2} .
\end{equation}
In order to verify this analytic estimate of the biased amplitude, we simulated full sky maps including a 
velocity dipole (with $v = 370\ \mathrm{km}\, \mathrm{s}^{-1}$). From these simulated catalogues 
we extracted the observed amplitude $d_{\mathrm{obs}}$. Figure \ref{3DGraph} shows the simulated data, 
the true value of the dipole amplitude and  a fit of the form $f(N) = \sqrt{D^2 + 9 A^2/N}$.

The best-fit values are $A = 0.908 \pm 0.002$ and $D = (0.451 \pm 0.001) \times 10^{-2}$ (statistical errors
only). These numbers should be compared to 
the factor $0.92$ from (\ref{estimateA}) and the simulation input of $d=0.46 \times 10^{-2}$. The reduced chi-
square of the fit is $7 \times 10^{-5}$. Thus  the Monte Carlo simulations agree with the 
theoretically motivated ansatz (\ref{ansatzd}) for the expected dipole amplitude of the estimator 
$\vec d_\mathrm{3D}$.

We conclude that even for a perfect full sky catalogue (no masking, complete in flux, perfect flux and 
position measurements), the amplitude of the linear estimator is biased towards higher values. 
Increasing the number of  radio sources will reduce the bias. The estimator $d_{\rm{3D}}$ is asymptotically  
unbiased, but this is of limited practical use for the analysis of NVSS and WENSS data.  
A similar bias of the dipole amplitude is also found for the other linear estimators introduced above. 
This finding is in agreement with \citet{Gibelyou}, who use a linear estimator to find the direction of the 
dipole.

\section{Linear estimators for an incomplete sky}

So far we assumed full coverage of the radio sky. More realistic catalogues cover just a fraction of the sky, 
as all earth based telescopes are limited to observe at certain declination ranges. 
Additionally, the Milky Way is masking parts of the sky. Here we will discuss some of the effects caused 
by incomplete sky coverage.
  
The upcoming Low Frequency Array (LOFAR) Tier-1 survey will cover about half of the sky ($2\pi$), thus we will first focus on this 
situation. As a second step we generalize this to an arbitrary axisymmetric survey geometry and include 
the effects of masking.
 
\subsection{Random walk}
\label{randomwalk}

Let us assume a survey geometry that covers all of the Northern hemisphere and ask how 
the estimator of the radio dipole (\ref{Nsum}) has to be modified. 
For the first two Cartesian components of $\vec{R}_\mathrm{3D}$ there should be no 
systematic problem, but the third component will definitely be biased. It is necessary to remove the effect of 
the incomplete sky from this $z$ component. Consider the expectation value of $\vec{R}_\mathrm{3D}$ for 
the Northern hemisphere

\begin{equation}
 \langle \vec{R}_\mathrm{3D} \rangle_{\mathrm{hemisphere}}=  \langle \sum_{i=1}^N \hat{r}_i \rangle=
\frac{N p_d}{2\pi}  \int_{\delta > 0} \mathrm{d}\Omega (1 + \hat{r} \cdot \vec{d}) \hat{r}, 
\end{equation}
where $p_d \equiv 1/[1 + (d/2) \cos \vartheta_d)]$ accounts for the proper normalisation of the probability
distribution on the hemisphere in presence of a dipole. $4\pi$ in (\ref{density2}) becomes $2\pi$ 
for obvious reasons.
The integral can be evaluated by hand. One finds
\begin{equation}
\langle \vec{R}_\mathrm{3D} \rangle_{\mathrm{hemisphere}}= N p_d \left(\begin{array}{c} \frac{1}{3} d
 \cos{\varphi}_d \sin{\vartheta}_d \\ \frac{1}{3} d \sin{\varphi}_d \sin{\vartheta}_d \\  \frac 12 + \frac{1}{3} d \cos
{\vartheta}_d\end{array}\right)  ,
\end{equation}
where $\vartheta_d$ and $\varphi_d$ denote the dipole position in spherical coordinates. 
The $z$ direction is strongly influenced by the
incomplete sky coverage. The total number of observed sources is not independent from the 
amplitude and orientation of the dipole itself. 

Nevertheless there is no problem in the evaluation of $\varphi_d$ because one can calculate
\begin{equation}
        \label{phiestimator}
 \varphi_d=\arctan{\frac{\vec{R}_y}{\vec{R}_x}}  .
\end{equation}
Here $N$ as well as $\sin{\vartheta}_d$ cancel out. So the 2D estimator is unbiased with respect to 
$\varphi_d$. Therefore we propose a pure two dimensional estimator:
\begin{equation}
\label{2Destimator}
 \vec{R}_{2D}=\sum_i^N \left( \begin{array}{c} \cos{\varphi}_i \sin{\vartheta}_i \\ \sin{\varphi}_i \sin{\vartheta}_i \\ 0 \end{array} \right)  .
\end{equation}
From this one can still use $(R_x^2+R_y^2)^{-1/2}$ for evaluating $d \sin{\vartheta_d} N p_d/3$ 
and $\varphi_d$. Let us take a look at the factor $\sin{\vartheta_d}$. Sources near the pole will make a 
smaller contribution than those further away. If a source near the pole is shifted by a small distance, the 
value $\varphi_i$ of this source could change dramatically. So the weighting terms compensate for 
this artificially big errors, which are just a relic of the coordinate system. 

Let us now estimate the uncertainties of the estimator $\vec R_{2D}$. The problem corresponds to 
an isotropic random walk process with variable step size. The probability density for a displacement 
of $r$ for such a random walk is
\begin{equation}
        \label{WalkResult}
 \check{P}_N(r)_{2D} \mathrm{d}r= \frac{3}{N} r \exp\left( -\frac{3r^2}{2N} \right) \mathrm{d}r  .
\end{equation}
 Similar to the random flight we determine $R_{pCL}$ defined by
\begin{equation}
 P_N(R_{2D}>R_{pCL})= \int_{R_{pCL}}^\infty \mathrm{d} r \check{P}_N(r)_{2D}=1- pCL  .
\end{equation}
Here again $pCL$ is the confidence level. 
It is possible to solve the above integral analytically
\begin{equation}
 \int_{R_{pCL}}^\infty \mathrm{d} r \frac{3}{N} r \exp\left( -\frac{3r^2}{2N} \right)  =
 \exp\left( -\frac{3R_{pCL}^2}{2N} \right)  .
\end{equation}
So $R_{pCL}$ is given by
\begin{equation}
        \label{erroresti1}
 R_{pCL} = \sqrt{\frac{2N}{3} \ln{ \left(\frac{1}{1-pCL} \right)}}  .
\end{equation}
Now one can use the same argument as for the random flight to evaluate the uncertainty of the 
$\varphi_d$ estimation
\begin{equation}
        \label{erroresti2}
 \delta \varphi_{pCL} = \arcsin{\frac{R_{pCL}}{R_{2D}}}  .
\end{equation}
In this way one can directly determine error bars for measured results of $\varphi_d$ calculated via 
(\ref{phiestimator}). Using this two dimensional estimator one cannot measure the dipole amplitude $d$ 
itself but only the combination $d \sin \vartheta_d N p_d/3$. Therefore it can only give an lower limit for $d$.

\begin{figure}[!ht]
\begin{center}
 \includegraphics[width=6cm,angle=270]{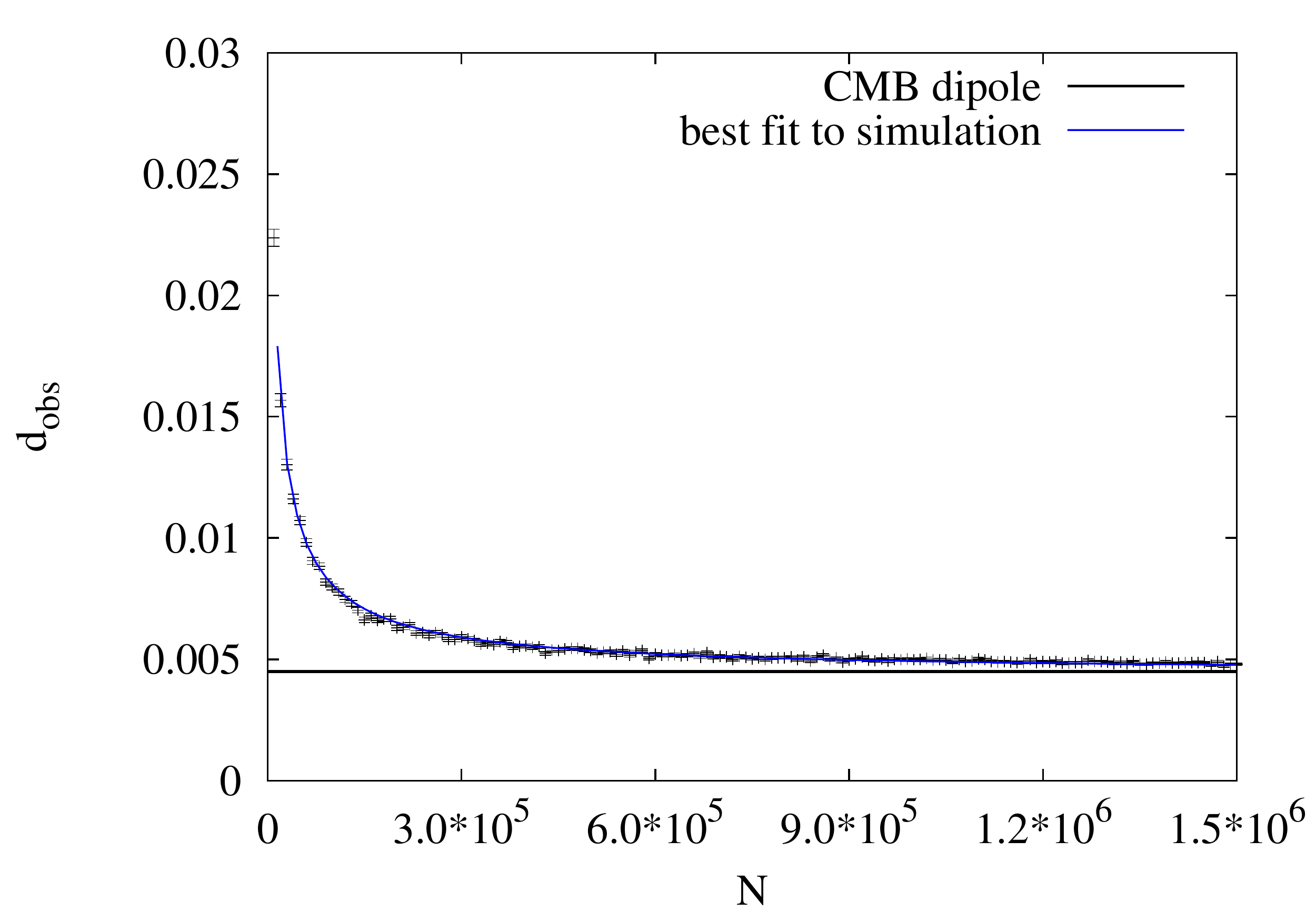}
\caption{
Ampliude bias for the estimator $d_{2D}$ on a hemisphere. Data represent mean and empirical 
variance of 1000 simulations for each N. A function $d_{\rm{obs}}(N) = \sqrt{D^2 + 9 A^2/N}$ 
is fitted to the simulated 
data, with best fit values $A= 0.712 \pm 0.003$ and $D= (0.444 \pm 0.002) \times 10^{-2}$. 
The  dipole amplitude ($0.0046$) is 
indicated by the horizontal line, the dipole vector is assumed to lie in the equatorial plane 
($\sin \vartheta_d = 1$).\label{2DGraph}}
\end{center}
\end{figure}

Like in the case of the full three dimensional estimator, this version is also biased in the measurement of 
the amplitude. The expectation of the random contribution can be calculated via 
\begin{equation}
 \langle R_{\mathrm{2D}}^{\mathrm{random}} \rangle= \int_0^\infty r   \frac{3}{N} r \exp\left( -\frac{3 r^2}{2N} \right) \mathrm{d}r  \approx 0.72 \sqrt{N} .
\end{equation}
So we expect our estimator to measure a combination of this random contribution and the true 
velocity dipole and make the ansatz
\begin{equation}
\langle d_{\mathrm{2D}}\rangle = \sqrt{ d^2 \sin^2 \vartheta_d  + \frac {9}{N^2} 
\langle R_\mathrm{2D}^\mathrm{random} \rangle^2}  ,
\end{equation}
where we used $p_d \approx 1$. Like above, we verify this via Monte Carlo simulations, shown in figure 
\ref{2DGraph}. 

\subsection{Direction bias}
\label{DirectionBias}

For any masked or incomplete map of the sky, 
we cannot measure the mean source density $\bar N/(4\pi)$, i.e.~the monopole. 
Therefore we always have to keep in mind that the observed mean density is just an approximation.
\citet{Gibelyou} have used an estimator proposed by \citet{Hirata}, which implicitly assumes, that the monopole is 
known. Based on the knowledge of the monopole, this estimator would compensate for 
masking effects and incomplete sky coverage by substracting a pure random isotropic map from 
the observed dipole term via (\ref{Gibelyouesti}). However, we cannot know $\bar N$.

For the previous example of a hemisphere, the monopole density is $\bar N/(4\pi) = N p_d/(2\pi)$, where 
$N = N_D$ in (\ref{Gibelyouesti}). 
The random map can be made up of an arbitrary number of sources $N_R$, but is reweighted by the 
observed number of sources $N_D/N_R$, instead of $\bar N/N_R$. This introduces a directional bias, 
in addition to the previously discussed biased amplitude. One can see this explicitly by applying 
(\ref{Gibelyouesti}) on our model of a dipole modified isotropic hemisphere. We find
\begin{equation}
\langle \vec{R}_\mathrm{3DM} \rangle_{\mathrm{hemisphere}}=
\frac{N p_d}{3}  d
\left(\begin{array}{c} 
 \cos{\varphi}_d \sin{\vartheta}_d \\ 
 \sin{\varphi}_d \sin{\vartheta}_d \\  
 \cos{\vartheta}_d\end{array}\right)  + \frac 1 2 N \left(\begin{array}{c} 
 0\\ 
 0 \\  
 p_d-1 \end{array}\right),
\end{equation}
which clearly is not parallel to $\vec d$. For small values of $d$ we can expand $p_d$ and obtain
\begin{equation}
 \langle \vec{R}_\mathrm{3DM} \rangle_{\mathrm{hemisphere}}=  \frac{N}{3}  d
\left(\begin{array}{c} 
 \cos{\varphi}_d \sin{\vartheta}_d \\ 
 \sin{\varphi}_d \sin{\vartheta}_d \\  
 \frac{1}{4} \cos
{\vartheta}_d\end{array}\right)  + O(d^2) .
\end{equation}
The $z$ component of the dipole is 
underestimated by a factor of $4$ for the geometry of a hemisphere. Despite cancelation of the 
leading term of the bias of the $z$ direction, the dipole direction remains biased. Less symmetric survey 
geometries lead to a bias of all dipole components for this estimator.

The best strategy to avoid any directional bias for a linear estimator is to make the survey 
geometry point symmetric around the observer for three dimensional estimators like 
$\vec R_{\mathrm{3D}}$ or point symmetric around the zenith in case of two dimensional estimators
$\vec R_{\mathrm{2D}}$. 
This implies for the NVSS that one has to cut symmetric in declination, such that both polar caps are missing, 
a strategy that was used by \citet{Singal11}.

\subsection{Masking}

The use of a masked sky additionally affects the dipole measurement. In general, the 
estimated dipole direction and amplitude depend on the position of the true dipole relative to the 
mask. Cutting areas with large dipole contribution will reduce the amplitude and vice versa.

In the following we consider masks that are point symmetric with respect to the observer
for all considered 3-dimensional estimators, respectively point symmetric with respect to the zenith for the 
2-dimensional estimators. Constructed in such a way, masking does not introduce any directional bias.
Nevertheless, we have to consider the effects of masking on the estimated dipole amplitude. 

A simple method to correct for this effect was put forward by \cite{Singal11}, 
who introduced a masking factor $k$,
\begin{equation}
\label{Nssum}
 \vec R_{\mathrm{3D}}^{\mathrm{mask}} = \frac{1}{k_\mathrm{3D}} \sum \vec r_i.
\end{equation}
Such a factor is expected to be a function of shape and position of the mask as well as the 
dipole position. In most cases an analytic calculation of this factor is impossible. 
For simple mask geometries some analytic results can be found in \citet{MatthiasThesis}. An alternative 
approach is to simulate the effects by means of Monte Carlos and to compare simulations with and without 
masking. The ratio of both results after a large number of simulations provides the masking factor $k_{\rm{3D}}$.

\begin{figure}[!ht]
\begin{center}
 \includegraphics[width=6cm,angle=270]{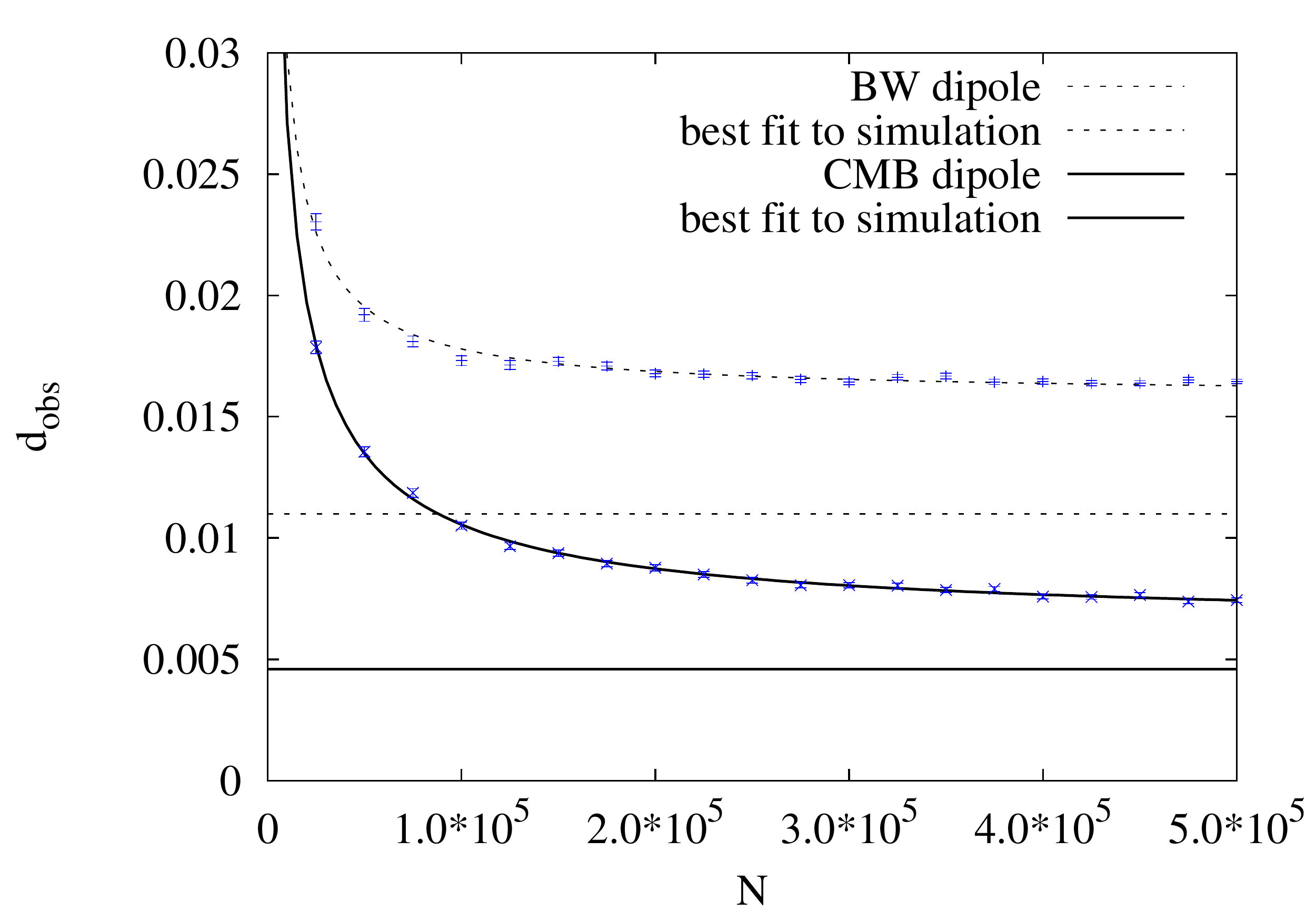}
\caption{Amplitude bias of the 3-dimensional estimator for the masked NVSS geometry of \citet{Singal11}. 
Data represent mean and empirical variance of 1000 simulations 
for each $N$. A function  $d_{\rm{obs}}(N) = \sqrt{(KD)^2 + 9 A^2/N}$ is fitted to the simulated data, 
with best-fit values $A= 0.883 \pm 0.006$, $KD= (0.642 \pm 0.005)\times 10^{-2}$ and 
$A=0.847 \pm 0.016$, $KD =(1.59 \pm 0.008)\times 10^{-2}$ for the expected kinetic radio dipole 
and the radio dipole measured by \cite{BW02}, respectively. The simulated dipole amplitudes, 
without masking, are indicated by the horizontal lines. \label{3dmaskinggraph}}
\end{center}
\end{figure}

Doing so, we found some interesting effects. The masking factors depend on the number of 
objects in the simulated maps as well as on the true dipole amplitude. This can be explained as follows.
The masking will mainly affect the kinetic dipole contribution, while the random dipole will increase due to  
the decrease of the number of objects in the masked catalogue. Therefore we expect the amplitude to be
\begin{equation}
\label{3Dmaskeq}
 \langle d_{\mathrm{3D}}\rangle_{\mathrm{mask}} = \sqrt{  k^2_\mathrm{3D}  d^2 + \frac{9}{N^2} \langle R_\mathrm{3D}^\mathrm{random} \rangle^2}  .
\end{equation}
$k_\mathrm{3D}$ should only depend on the properties of the mask and not on the amplitude of the dipole. 
To test this, we created simulated maps for two different dipole amplitudes, with values 
motivated by the CMB measurement $d_{\mathrm{CMB}}=0.46 \times 10^{-2}$ and by 
the measurement of \citet{BW02} at $25$ mJy $d_{\mathrm{BW}}=1.1 \times 10^{-2}$.
We used the direction $\mathrm{RA}=158^\circ, \mathrm{dec}=-4^\circ$ in both cases. 
The mask agrees with the one used by \citet{Singal11}, i.e. we removed all sources  with 
$|\delta| > 40^\circ$ and $|b| < 10^\circ$. Resulting amplitudes for different numbers of sources are shown 
in figure \ref{3dmaskinggraph}.

First of all we can conclude that (\ref{3Dmaskeq}) is a good fit for the behaviour of the simulated maps in 
both cases. The measured amplitudes are larger than those estimated from full sky maps. 
We can now calculate $k_{\rm{3D}}$.  In both cases it turns out to be $1.4$, which could be used 
to correct the amplitude estimate. 

\begin{figure}[!ht]
\begin{center}
 \includegraphics[width=6cm,angle=270]{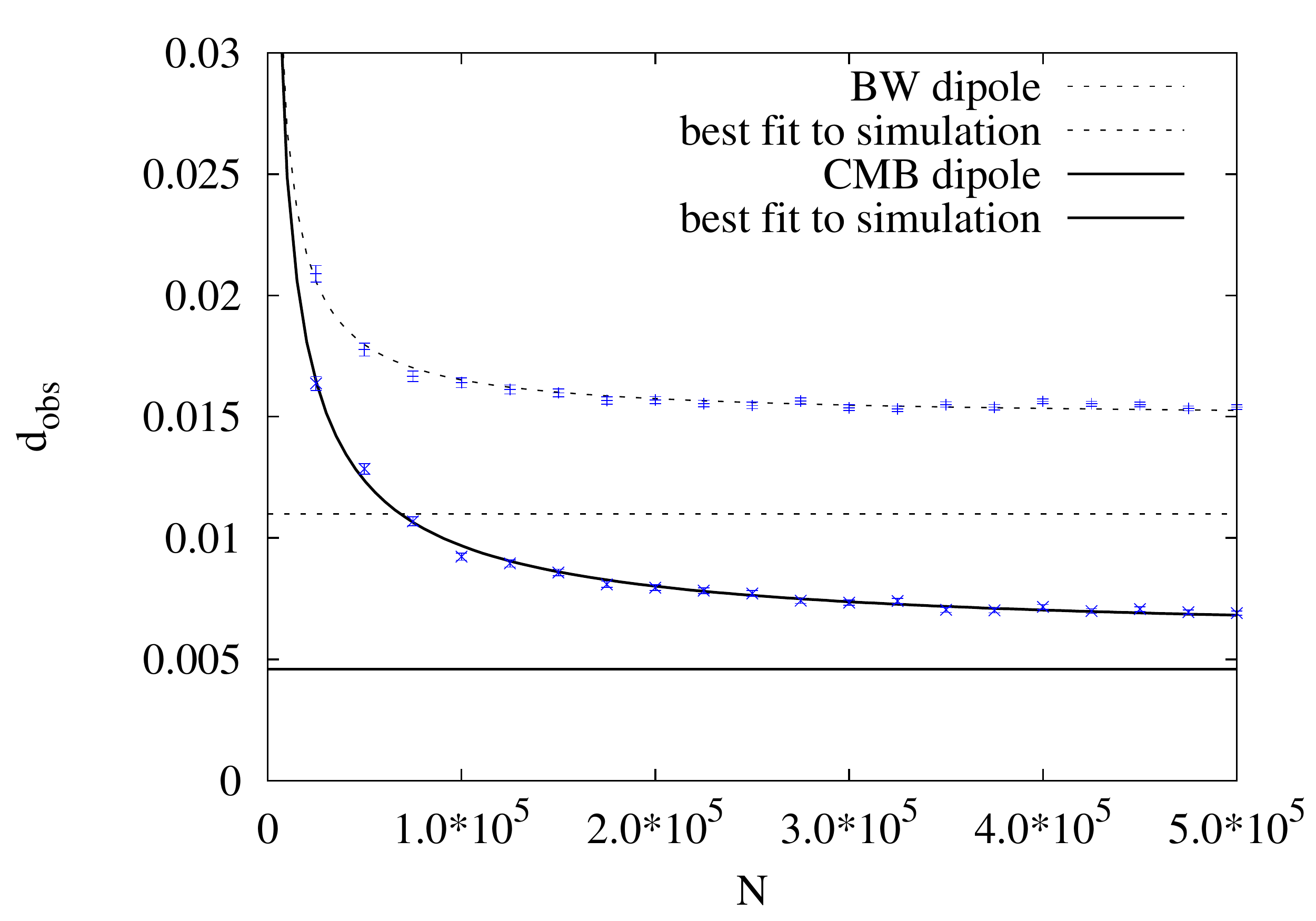}
\caption{Amplitude bias of the 2-dimensional estimator for the masked NVSS geometry of \citet{Singal11}. 
Data represent mean and empirical variance of 1000 simulations 
for each $N$. A function $d_{\rm{obs}}(N) = \sqrt{(KD)^2 + 9 A^2/N}$ is fitted to the simulated data, 
with best-fit values $A= 0.810 \pm 0.009$, $KD= (0.589 \pm 0.008)\times 10^{-2}$ and 
$A=0.745 \pm 0.014$, $KD =(1.493 \pm 0.006)\times 10^{-2}$, for the expected kinetic radio dipole 
and the dipole measured by \citet{BW02}, respectively. The simulated dipole amplitudes, without 
masking, are indicated by the horizontal lines. \label{2demaskfig}}
\end{center}
\end{figure}

A similar argument holds for the two dimensional estimator. Here we expect a behaviour of the form:
\begin{equation}
 \langle d_{\mathrm{2D}}\rangle_{\mathrm{mask}} = \sqrt{  k^2_{2D} d^2 \sin^2\vartheta_d+ 
 \frac{9}{N^2}\langle R_\mathrm{2D}^\mathrm{random} \rangle^2}  .
\end{equation}
Using the same assumptions about the dipole term and the same mask as before, we obtain 
figure \ref{2demaskfig}.

This time, we find $k_{2\mathrm{D}}=1.3$ for both assumed velocities. The simulations 
support our assumption that the masking factor does not depend on the dipole magnitude $d$. 

\subsubsection{Masking correction}
\label{MfS}

Although the masking factor $k$ does not depend on the amplitude $d$, it may depend on the dipole 
direction $\hat{\vec{d}}$. 
Therefore it would be necessary to repeat the analysis of the previous section for each 
dipole direction found. In order to reduce the simulation effort, we rely on the following method instead. 

For the full, as well as for the masked sky, surveys with $10^6$ sources were simulated. This 
choice guarantees that we investigate masking effects and not effects due to shot noise.
The mean dipole amplitudes are determined for $10^3$ simulated full and masked sky surveys,
respectively. The ratio of the masked sky mean amplitude to the full sky mean amplitude is  denoted 
$\tilde k $. This ratio provides a first approximation to the masking factor.
\begin{equation}
   \tilde k  \equiv \frac{\sqrt{  k^2  d^2 + \frac{9}{N^2} \langle R^\mathrm{random} \rangle^2} }
  {\sqrt{  d^2 + \frac{9}{N^2} \langle R^\mathrm{random} \rangle^2}}.
\end{equation}
The influence of the random dipole tends to bias $\tilde k$ towards $1$ (as can be easily seen in the limit of a small number of sources). This bias can be compensated by rewriting the above formula into
\begin{equation}
 k=\sqrt{ \tilde  k^2 + 
 \frac{9 \langle R^\mathrm{random} \rangle^2}{d^2N^2}\left( \tilde k^2-1 \right)}.
\end{equation}
The values of $N$ and $d$ are input parameters of the simulation. From the last section we know 
the values of $ \langle R^\mathrm{random} \rangle$ for the three- as well as for the two-dimensional 
case. Therefore we can transform the approximated masking factor $ \tilde k $ 
into the unbiased masking factor $k$.

\section{Dipole from a flux weighted estimator} 
\label{xProblem}

Let us now turn to a discussion of the flux weighted estimator (\ref{Ssum}), which was
used by \citet{Singal11}.

We first take a closer look at the theoretically expected value of the flux dipole $\vec d _\mathrm{flux}$. 
For simplicity we assume full sky coverage. For a large number of sources,
\begin{equation}
 \vec{\hat{d}} \cdot \sum_{i=1}^N S_i \vec{r}_i  \approx 
 \int_{4\pi}  \mathrm{d}\Omega 
 \int_{S_\mathrm{min}}^{S_\mathrm{max}} \mathrm{d}S
 \frac{\mathrm{d}^2N}{\mathrm{d} \Omega \mathrm{d}S}  S \cos\theta,
\end{equation}
where $\theta$ is the angle between a source and the dipole direction $\vec{\hat{d}}$. 

We now determine the number of sources per flux and solid angle as a function of the observer velocity. 
At zeroth order in velocity, this density $n(S)$ is isotropic. As for the number counts, 
stellar aberration and the Doppler effect have to be taken into account. Stellar aberration gives rise to
\begin{equation}
\frac{\mathrm{d}^2N}{\mathrm{d} \Omega \mathrm{d}S}  \approx n(S) (1+ 2 \frac{v}{c} \cos \theta).
\end{equation}
The relativistic Doppler effect alters the observed fluxes. When we observe a source in the 
direction of motion, we measure a higher flux than if we were at rest with respect to the isotropic 
and homogeneous Universe. We relate the observed flux $S_0$ to the flux that is 
measured by an observer with vanishing peculiar motion,
\begin{equation}
S_{\rm{rest}} \approx S_0 - S_0(1+\alpha)\frac{v}{c}\cos \theta .
\end{equation}
We assume the power law $n(S) = a S^{-\tilde x}$ to hold for observers at rest and Taylor expand around
the observed flux
\begin{equation}
n(S_{\rm rest}) \approx  n(S_0)[1 +  \tilde x (1+\alpha)\frac{v}{c} \cos \theta ] .
\end{equation}
If the assumed power law holds for all sources of a survey, then $\tilde x = 1 + x$, with $x$ as defined 
above in (\ref{numbercountpowerlaw}).
Combining the Doppler effect and the stellar aberration leads to
\begin{equation}
 \frac{\mathrm{d}^2N}{\mathrm{d} \Omega \mathrm{d}S} 
 =n_0(S_0)[1+(2+\tilde x (1+\alpha))\frac{v}{c}\cos \theta] + O \left( \left( \frac{v}{c}\right)^2 \right) .
\end{equation}
This result only holds under the assumption of a power law behaviour of the number counts. It is crucial to 
keep this in mind. 

\subsection{NVSS}

\begin{figure}[!ht]
\begin{center}
\includegraphics[width=6cm,angle=270]{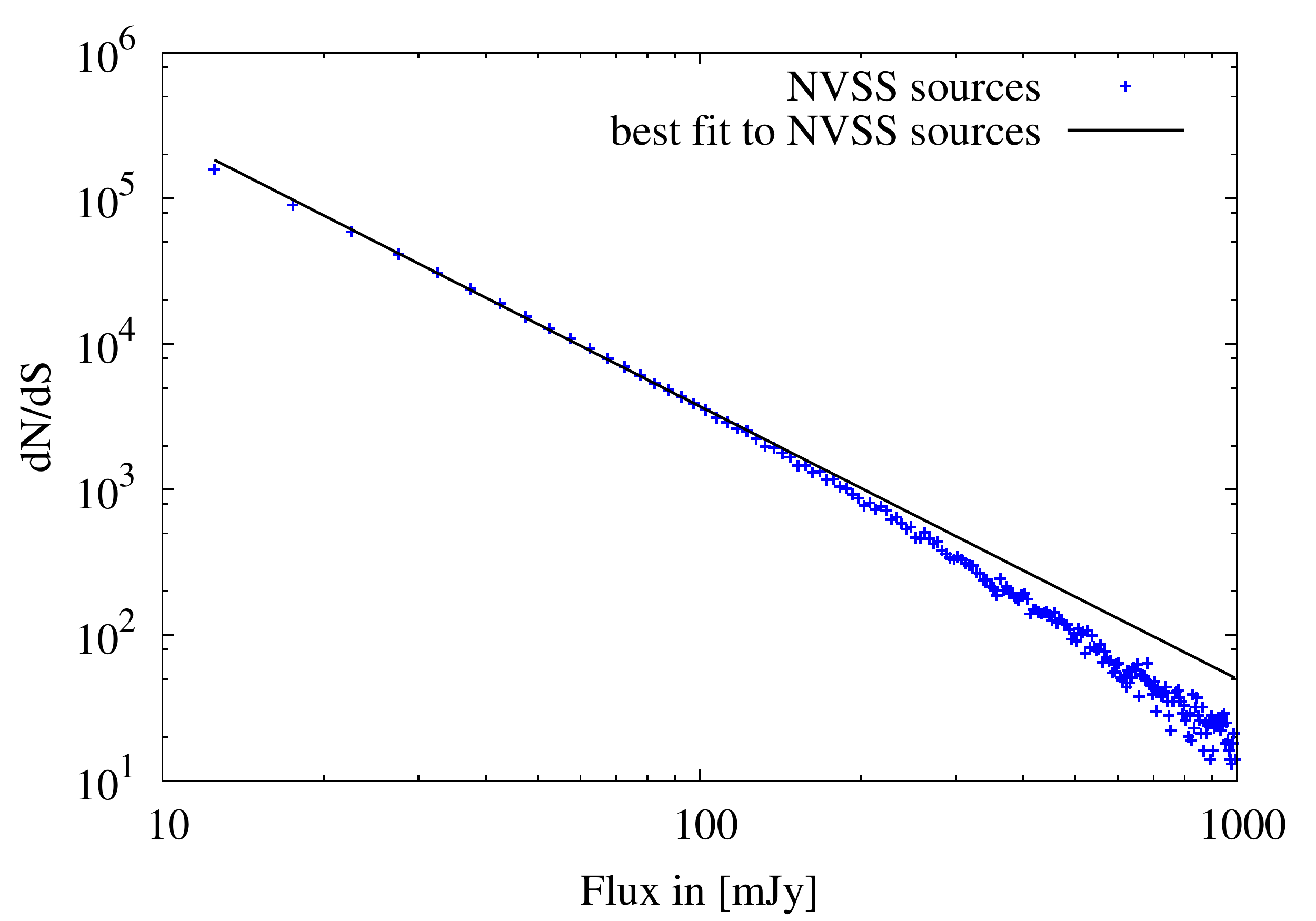}
\caption{Differential number counts of the NVSS catalogue, $S_{min}=10$~mJy, best fit values for 
$f(s)=a \cdot s^{-\tilde x}$ with $25\ \mathrm{mJy} < S < 1000\ \mathrm{mJy}$ 
are $a=2.1 \times 10^7$ and $\tilde x=1.9$.}\label{NVSSnvonS}
\end{center}
\end{figure}

Let us now see, if we are allowed to make this assumption for the analysis of NVSS data. 
The plot in figure \ref{NVSSnvonS} demonstrates that the power law is not valid for the flux range
used in the analysis of \citet{Singal11}, i.e.~fluxes up to 1 Jy.
Actually the slope steepens for the larger fluxes considered. 
The best fit power-law gives a reduced chi-squared value of $\chi^2=122$.
That means that the observed data cannot be fitted by a power law. 

Thus Singal's assumption $\tilde x \approx 1$ does not hold for two reasons. Firstly, for a pure power law 
we would expect $\tilde x = 1 + x \approx 2$, which is close to what we find: $\tilde x \approx 1.9$.
Secondly, the power-law assumption only applies to about one half of the data. 

In conclusion, the result of unexpectedly large amplitudes in \cite{Singal11} could 
partly be explained by this two effects. With power law behaviour one should use 
$\tilde x=1.9$ or larger to also account for the 
steepening of the spectral index at high fluxes, which increases the expectation value by
\begin{equation}
 \frac{\langle \vec d_{\mathrm{obs}}\rangle|_{\tilde x=1.9}}{ \langle\vec d_{\mathrm{obs}} \rangle|_{x=1}} 
 > 1.4 . 
\end{equation}
The results of the flux weighted estimator from \cite{Singal11} should be reduced by at least a 
factor of $1.4$. 

Compared to the estimator of Crawford, the estimator (\ref{Ssum}) stresses sources with high flux. 
To avoid the domination of a small number of sources, sources with $S > 1000$ mJy are not taken 
into account. Nevertheless, this estimator is stressing bright sources.  
These sources are, on average, nearer than the rest. Hence (\ref{Ssum}) might be dominated by 
nearby sources and by atypically bright ones. This seems to be yet another weakness of this estimator, 
since the local universe is anisotropic. 

\section{Dipole estimates from NVSS and WENSS}
\label{Ourresults}

\subsection{3D linear estimates} 

We are now in a position to check the three dimensional estimations of the radio dipole in the NVSS catalogue. 
As we have shown above, the estimator used by \citet{Gibelyou} (\ref{Gibelyouesti}) gives rise to a biased dipole direction and 
thus is not further considered in this work. The flux weighted estimator (\ref{Ssum}) is also of limited use, as the NVSS 
data cannot be described by a power-law over all fluxes of interest. We thus focus here on the simplest 
linear estimator (\ref{Nsum}).  

In order to obtain an unbiased direction estimate, the cut sky geometry of \cite{Singal11} is adopted. 
The masking factor $k$ is determined for every measured dipole anisotropy direction, as described 
in \ref{MfS}. The pure estimator results $d_{3D}$ are then corrected for the masking bias and we 
obtain $d_{3D}^{\mathrm{cor}}$. 

\begin{table}[ht!]
\begin{tabular}{ccccccc}
\hline
Flux&  N &  RA & dec & $d_{3D}$ & $k$ & $d_{3D}^{\mathrm{cor}} $\\ 
(mJy)  & & ($^{\circ}$)&($^{\circ}$)  & ($10^{-2}$) & &($10^{-2}$) \\ \hline
   50   & $ 91,662 $& $ 170 \pm  23 $ &$ -17\pm  23 $& $ 2.78 $ & $1.38$&  $ 2.0 \pm 0.8$ \\ 
   40    & $ 115,917 $& $ 156 \pm  26 $ &$ -18 \pm  26 $& $  2.23 $& $1.29$& $ 1.7 \pm 0.8$ \\ 
   35  & $ 133,022 $ & $ 156 \pm  22 $  &$ -11 \pm  22 $& $  2.46 $& $1.32$& $ 1.9 \pm 0.7$ \\ 
   30  & $ 155,120 $ & $ 156 \pm  19 $  &$ -2 \pm 19 $& $  2.63 $& $1.35$& $ 1.9 \pm 0.7$ \\ 
   25   & $ 185,649 $ & $ 158 \pm  19 $ &$ -2 \pm  19 $& $  2.38 $& $1.34$& $ 1.8 \pm 0.6$ \\ 
   20   & $ 229,557 $& $ 153  \pm  18 $ &$ 2 \pm  18 $& $ 2.31 $& $1.30$& $ 1.8 \pm 0.6$ \\ 
   15  & $ 298,289 $ & $ 149 \pm  18 $ &$ 17 \pm  18 $& $ 2.02 $& $1.28$ & $ 1.6 \pm 0.5$ \\ 

\hline
\end{tabular}
\caption{Dipole direction and amplitude from NVSS. The estimator (\ref{Nsum}) was used. Excluded are sources with $|\delta|>40^\circ$ and $|b|<10^\circ$ (J2000).\label{ReproSingalresults}}
\end{table}

All results for right ascension agree within their error bars. The same holds true 
for the dipole amplitudes. This self-consistency indicates the absence of significant 
systematic errors. Unfortunately, we can not state the same for the declination results. 
One observes an significant increase in declination with respect to deceasing flux limits. 
This effect is very likely due to a relic of the NVSS survey procedure. Sources below 
$\delta=-10^\circ$ were measured by means of a different alignment of the Very Large Array. 
The source density is therefore smaller in this area. Therefore it is expected that the dipole 
measurement will show increasing values of declination at the fainter flux limits. This makes 
it hard to trust the declination results below a flux limit of $25$ mJy, (note that there is no 
significant difference between 20 mJy and 30 mJy).

The results of table \ref{ReproSingalresults} can be compared to those in table \ref{SingalresultsN}. 
Number of sources and direction results are almost the same. Deviations could be explained by 
minor differences in the precise form of the mask. We used a different method for estimating 
uncertainties in the direction and amplitude measurement, than those used by \citet{Singal11}. 
Since our method (described above) is more conservative, we obtain larger errorbars. All dipole 
amplitudes in this table are slightly below those from table \ref{SingalresultsN}. In \citet{Singal11} a 
different method was used to obtain the dipole amplitude from the estimator (\ref{Nsum}), which can 
explain this discrepancy. In principle we can recover the results of \citet{Singal11} and confirm an 
unexpectedly high dipole amplitude.

\subsection{2D linear estimates}

A disadvantage of the NVSS catalogue is that the sampling depth changes at $\delta = -10^\circ$ 
(less sensitivity at lower declinations).  This could lead to a directional bias of the NVSS data analysis and 
thus it is interesting to also use the two dimensional estimator presented in this work, as this effect cannot 
lead to a bias in this case. For the WENSS analysis, a three dimensional linear estimator cannot avoid 
directional bias. 

\subsubsection{NVSS}

A major advantage of the estimator $\vec R _{2D}$ is that it does not require a north-south symmetry 
of the catalogue. Therefore, the declination limit of the NVSS catalogue is no problem.

\begin{figure}[!ht]
\begin{center}
 \includegraphics[width=\linewidth]{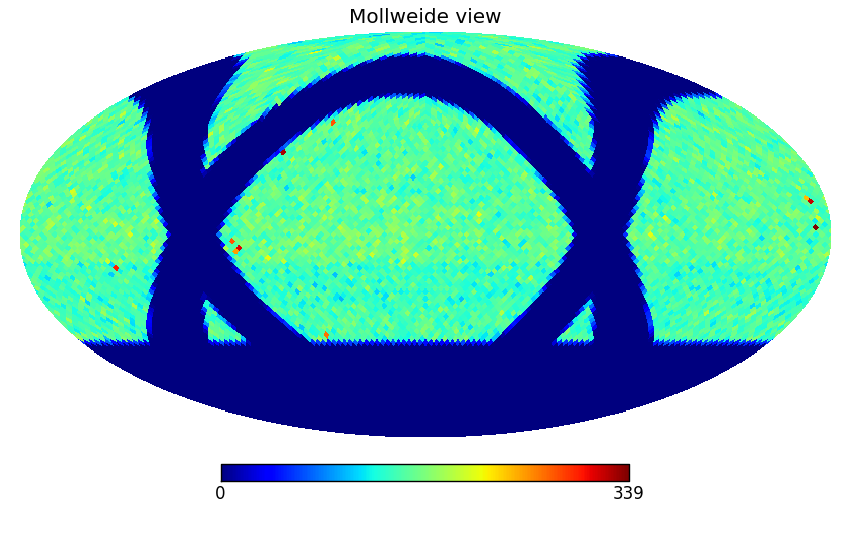}
\caption{A map of the number counts in HEALPix pixels from NVSS. The pixel size corresponds to  
$N_{\rm side} = 32$. Shown are equatorial coordinates at epoch J2000. 
The NVSS contains data at $\delta > -40^\circ$ and the galactic plane and a ``counter galaxy'' are masked 
(CG mask) in order to avoid galactic contamination and to 
restore point symmetry with respect to the zenith.}\label{NVSScut}
\end{center}
\end{figure}

As the estimator $\vec R_{2D}$ requires a point symmetry around the north pole, we cannot remove the 
Galactic plane only. For each removed point we also need to subtract the point which is $180^\circ$ away. 
When we do so, a second plane occurs, which we call the Counter Galactic plane (CG). A HEALPix\footnote{http://healpix.jpl.nasa.gov} map of 
the remaining NVSS sources can be seen in figure \ref{NVSScut}. The color of the pixels encodes the 
number of sources per pixel. 

An alternative would be the mask used by \citet{Singal11}. That mask is a combination of two 
great cycles and would therefore also work for $\vec R_{2D}$. However, it turns out that the 
mask with the CG removes fewer sources. Thus we decided to use the CG mask.

The next step is to evaluate the masking correction $k$ of the CG mask. As this factor depends on 
the right ascension and on the declination of the dipole anisotropy, we need some additional 
information. From $\vec R_{2D}$ we estimate the right ascension. As the declination cannot be 
evaluated with $\vec R_{2D}$, we use the declination as provided by the three dimensional estimator
in order to determine $k$. These values are certainly not exact, since a different 
mask is used now. The influence of a small change in dipole declination on the evaluated factor 
$k$ is discussed in section \ref{WENSSsection}.

We reduce the dipole amplitude of the 2D estimator by the masking factor $k$ and 
obtain the masking corrected amplitude $d_{\mathrm{2D}}^{\mathrm{cor}}$.
The results of this procedure for the NVSS catalogue and different flux limits can be found in table
\ref{NVSSresults}.

\begin{table}[ht!]
\begin{tabular}{cccccc}
\hline
Flux&  N &  RA  & $d_{2D} $ & $k$ & $d_{2D}^{\mathrm{cor}} $\\ 
(mJy)  & & ($^{\circ}$) & ($10^{-2}$) & &($10^{-2}$) \\ \hline
   50   & $ 96,337 $& $ 171 \pm 19 $ & $ 2.63 $ & $1.42$&  $1.9 \pm 0.7$ \\ 
   40    & $ 121,831 $& $ 146 \pm 20 $ & $ 2.28 $& $1.29$& $1.8 \pm 0.7$ \\ 
   35  & $ 139,851 $ & $ 152 \pm 17 $  & $ 2.49 $& $1.28$& $1.9 \pm 0.6$ \\ 
   30  & $ 163,208 $ & $ 153 \pm 15 $  & $ 2.55 $& $1.28$& $2.0 \pm 0.6$ \\ 
   25   & $ 195,245 $ & $ 155 \pm 14 $ & $ 2.45 $& $1.29$& $1.9 \pm 0.5$ \\ 
   20   & $ 241,399 $& $ 150  \pm 14 $ & $ 2.25 $& $1.26$& $1.8 \pm 0.5$ \\ 
   15  & $ 313,724 $ & $ 148 \pm 15 $ & $ 1.86 $& $1.21$ & $1.5 \pm 0.4$ \\ 
   10    & $ 447,459 $& $ 133 \pm 14 $ & $ 1.67 $& $1.10$ & $1.5 \pm 0.4$ \\  
\hline
\end{tabular}
\caption{
Dipole right ascension and amplitude $d \sin \theta_d$ from NVSS. 
The 2D estimator (\ref{2Destimator}) is used and all sources with $\delta>-40^\circ$, except the galactic and counter galactic planes (CG mask), are included.
\label{NVSSresults}}
\end{table}

Obtained right ascensions and amplitudes are stable with respect to different flux limits. For 
all flux limits $\geq 15$ mJy, the estimated right ascension of the radio dipole is in agreement with the 
CMB prediction of $RA=168^\circ$. Only when we include sources as faint as $10$ mJy, find a 
$3\sigma$ deviation. However, we know that the catalogue is incomplete at its faint end.

The masking corrected dipole amplitudes $d_{2D}^{\mathrm{cor}} $ are significantly above the 
CMB prediction. To some extent this is expected due to the discussed amplitude bias. A detailed 
discussion is presented in the next section.

\subsubsection{WENSS}
\label{WENSSsection}

\begin{figure}[!ht]
\begin{center}
 \includegraphics[width=\linewidth]{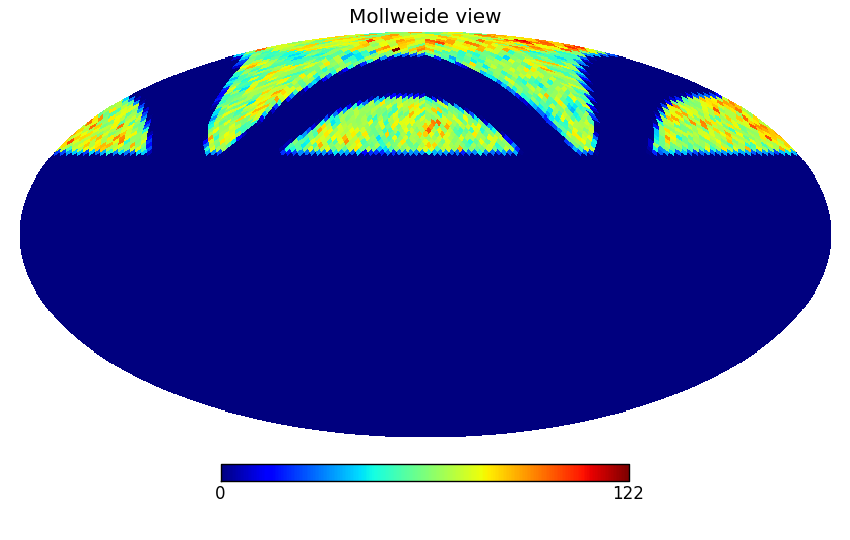}
\caption{A map of the number counts in HEALPix pixels from WENSS. The pixel size corresponds to  
$N_{\rm side} = 32$. Shown are equatorial coordinates at epoch B1950. 
The WENSS contains data at $\delta > 30^\circ$ and the galactic plane and a ``counter galaxy'' are masked 
(CG mask) in order to avoid galactic contamination and to 
restore point symmetry with respect to the zenith.}\label{WENSScut}
\end{center}
\end{figure}

We finally present the first estimation of the radio dipole from the WENSS catalogue. 
We cannot use the three dimensional linear estimators on this catalogue, since it only contains sources with 
$\mathrm{dec}>28^\circ$. The two dimensional estimator on the other hand can be used here. 
Again, we need to remove the Galactic plane and the Counter Galactic plane to reestablish a 
symmetry around the north pole. The remaining WENSS catalogue is shown in figure \ref{WENSScut}.

\begin{figure}[!ht]
\begin{center}
 \includegraphics[width=6cm,angle=270]{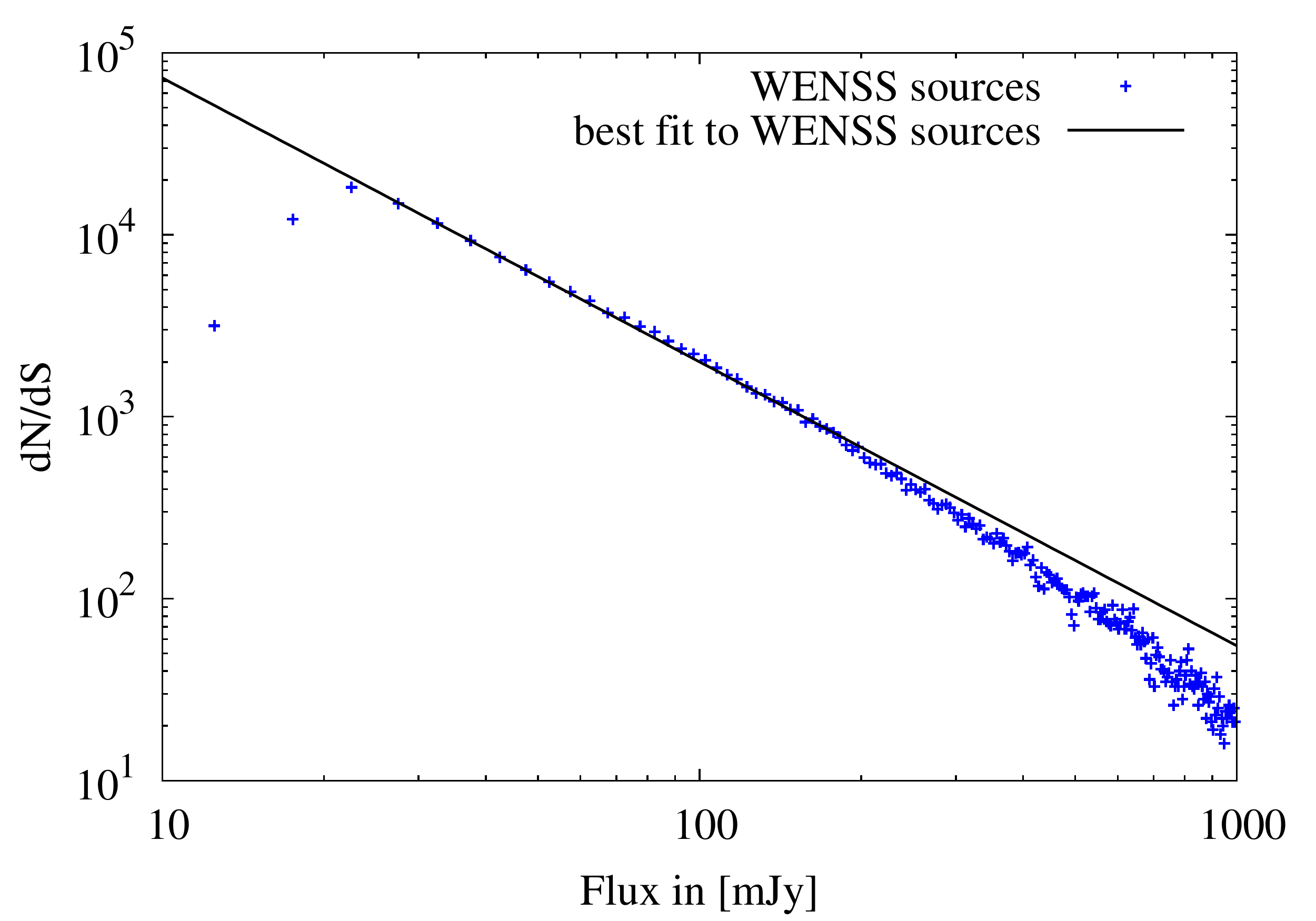}
\caption{Differential number counts of the WENSS catalogue, $S_{min}=5$~mJy, best fit values for $f(s)=a \cdot s^{-\tilde x}$ with $25\ \mathrm{mJy} < S < 1000\ \mathrm{mJy}$ are $a=2.6 \times 10^6$ and $\tilde x=1.6$.}\label{WENSSNvonS}
\end{center}
\end{figure}

In order to choose the best flux limit, we analyse the differential number counts between 
$10$ mJy and $1000$ mJy. Figure \ref{WENSSNvonS} shows that the WENSS catalogue 
seems to be incomplete below $25$ mJy. In \citet{WENSS} the completeness of the WENSS 
catalogue is claimed to hold only above a limit of $30$ mJy. From figure \ref{WENSSNvonS} 
we infer that a simple power law also applies to the source counts between $25$ and $30$ mJy and thus 
we include sources down to a flux limit of $25$ mJy in our analysis.

We cannot obtain information on the declination of the dipole from the WENSS catalogue
by means of the two dimensional estimator applied in this work. 
This could in principle be a problem for the determination of the masking factor $k$. Therefore, we 
further investigated the effect of different dipole declinations. Assuming the WENSS symmetry and a 
right ascension of $120^\circ$ (close to the results given in table \ref{WENSSfinal}), we calculated 
$k$ for $7$ different values of declination, see Table \ref{WENSSk}. For this mask, the dependence of $k$ on 
the right ascension of the dipole is relatively small, compared to shot noise uncertainties. 
We assume $\mathrm{dec}=0^\circ$ for the determination of $k$, 
based on the expectation from the CMB dipole and the NVSS radio dipole estimates.

\begin{center}
\begin{table}[h]
\begin{tabular}[h!]{cccccccc}
\hline
dec & $45^\circ$ & $30^\circ$ & $5^\circ$ & $0^\circ$ & $-5^\circ$ & $-30^\circ$ & $-45^\circ$ \\
\hline
$k$ & $0.50$ & $0.48$ & $0.45$ & $0.44$ & $0.44$ & $0.47$ & $0.51$\\
\hline
\end{tabular}
\caption{Masking correction $k$ for WENSS with CG mask and a dipole with $RA=120^\circ$.
\label{WENSSk}}
\end{table}
\end{center}

\begin{center}
\begin{table}[h]
\begin{tabular}[h!]{cccccc}
\hline
 Flux &  $N$&  RA  & $d_{2D}$& $k$ & $d_{2D}^{\mathrm{cor}} $\\
 (mJy) & & ($^{\circ}$)  & ($10^{-2}$)  & &($10^{-2}$)\\ \hline
   $40$  & $67,052$& $ 124 \pm 51 $ & $1.31$&$0.45$& $2.9 \pm 2.3$\\ 
   $35$  & $73,653$ & $ 123 \pm 46 $& $1.36$&$0.47$& $2.9 \pm 2.1$\\ 
   $30$  & $81,863$ & $ 122 \pm 48 $& $1.24$& $0.45$ & $2.8 \pm 2.1$\\ 
   $25$ & $92,600$& $ 117 \pm 40 $  & $1.36$&$0.47$& $2.9 \pm 1.9$\\   
\hline
\end{tabular}
\caption{Dipole estimate from WENSS based on 2D estimator using peak flux values for all sources with 
$\delta>30^\circ$, except those in the galactic and counter galactic planes (CG mask). Our WENSS 
analysis uses positions at epoch B1950.
\label{WENSSfinal}
}
\end{table}
\end{center}

The results of the WENSS analysis are presented in Table \ref{WENSSfinal}. 
Although the WENSS catalogue covers only one fourth of the sky, we find that it can be used for the 
estimation of the radio dipole. A problem is the limited number of sources that are left after masking the 
galaxy and restoring the required symmetry of the catalogue. This leads to larger error bars, compared with 
the NVSS estimates. We can conclude that the observed dipole anisotropy in the WENSS catalogue is 
in agreement with the measurements from NVSS, which is a nontrivial statement, as we are probing 
radio sources at different frequencies ($325$ MHz vs.~$1.4$ GHz).

\section{Comparison of results} 
\label{Discussion}
\begin{figure*}[ht!]
 \includegraphics[width=6cm,angle=270]{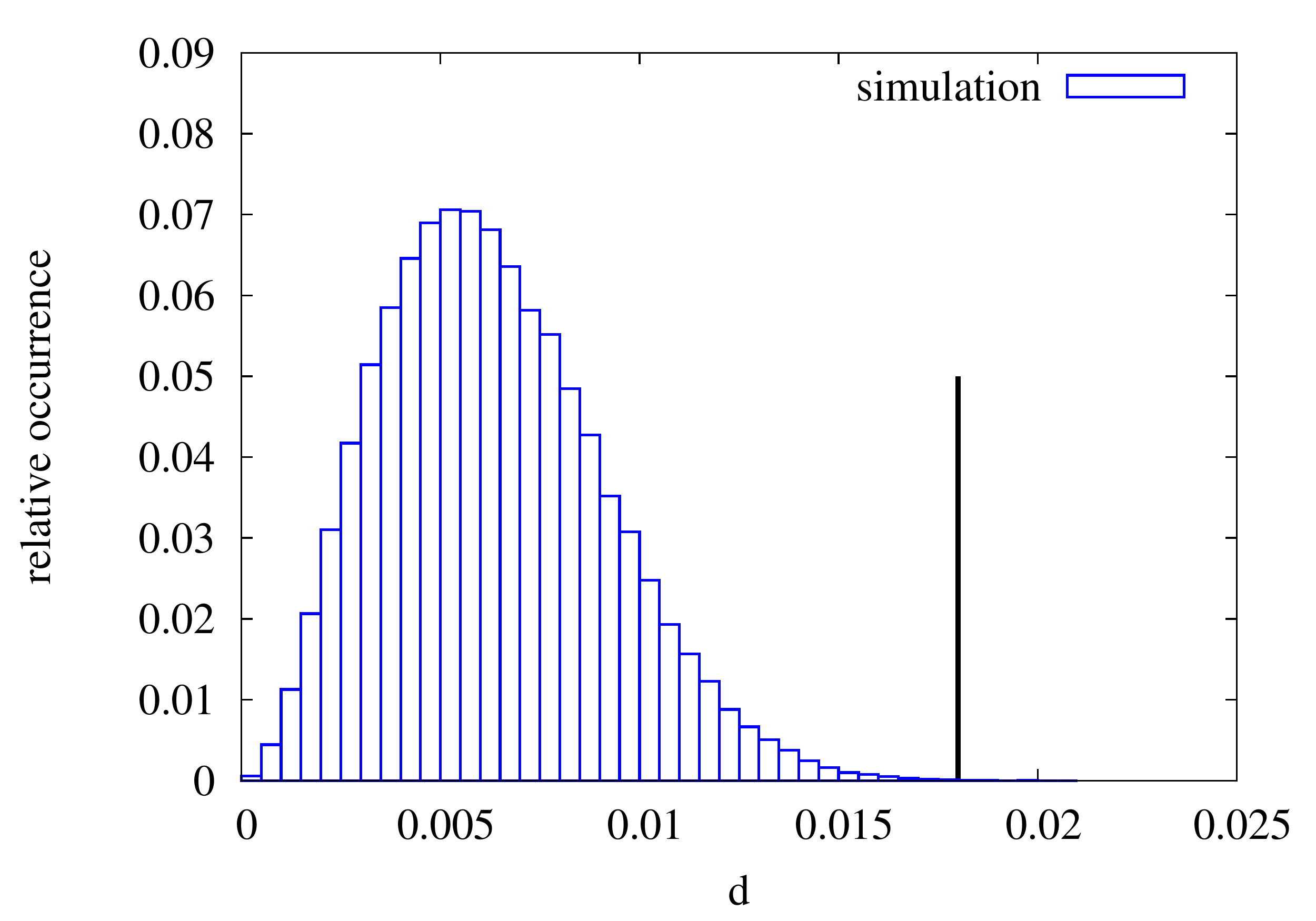} \hspace{0.6cm}
 \includegraphics[width=6cm,angle=270]{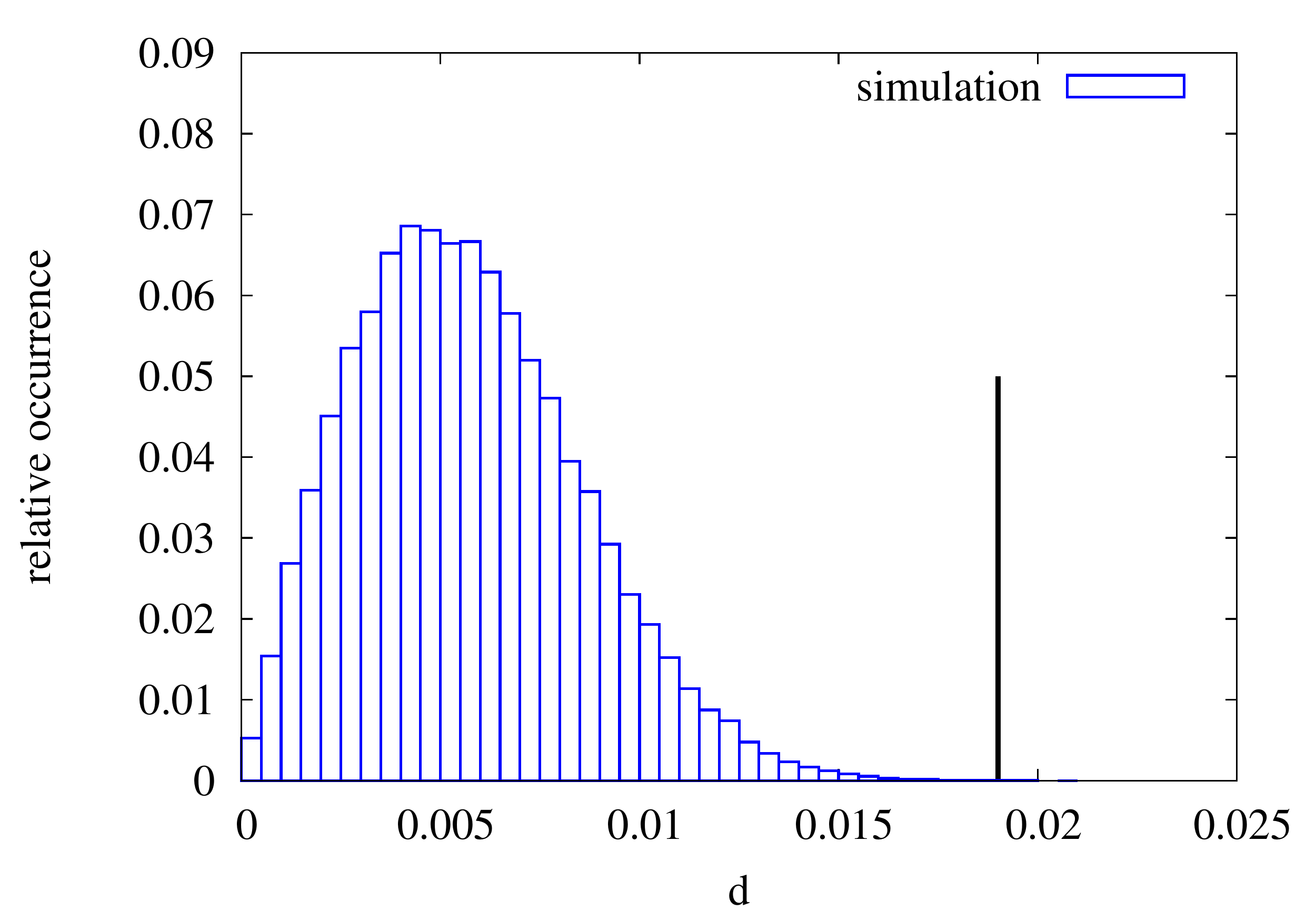}
\caption{Histogram of dipole amplitudes for 100,000 simulations of the three dimensional (left) and 
two dimensional (right) estimator, assuming the CMB expectation and a slope of $x=1.1$, 
with 185,649 (left) and 195,245 (right) sources per simulation and appropriate masks. 
The black vertical lines are the NVSS results of this work.\label{HistoBoth}}
\end{figure*}

We summarize the various results from the literature and this work in table \ref{Comparison}. 
The results of this work are highlighted with bold faced letters. For comparison we focus on the 
flux limits of $25$ mJy and $15$ mJy. 

All estimated dipole directions, both from the NVSS and from WENSS are in 
good agreement with each other and with the direction from the CMB dipole, with the exception 
of the result from \citet{Gibelyou}. As explained in section \ref{DirectionBias}, their estimator shows a 
directional bias. We did not investigate any further, whether this bias invalidates their 
findings at a rather low flux limit. Our analysis based on the three dimensional estimator applied to 
NVSS and using the mask defined by \citet{Singal11} gives (RA, dec) $= 
(158^\circ \pm  19^\circ, -2^\circ \pm  19^\circ)$.

\begin{table}[ht!]
\begin{tabular}{cccccc}
\hline
Source & Flux $>$ & N&   RA & dec  & $d$\\ 
 &(mJy) & & ($^{\circ}$)  & ($^{\circ}$)  & ($10^{-2}$) \\ \hline
{\bf  NVSS} & & & & & \\ \hline
   BW  & $ 25 $ & $ 197,998 $ & $ 158 \pm 30 $ & $-4 \pm 34$&  $1.1 \pm 0.3$ \\ 
   SIN  & $ 25 $  & $ 185,474 $ & $ 158 \pm 10 $& $-2\pm 10$& $1.8 \pm 0.4$ \\ 
   SIF  & $ 25 $  & $ 184,237 $ & $ 159 \pm 10 $& $-7\pm 9$& $2.2 \pm 0.6$ \\ 
   \bf SIF*  & 25  & $184,237$ &$ 159 \pm 10 $ &$-7\pm 9$& ${\bf 1.6 \pm 0.5}$ \\ 
{\bf 3DS} & $\bf 25$   & $\bf 185,649 $ & ${\bf 158 \pm 19}$ & ${\bf -2 \pm 19}$ & ${\bf 1.8 \pm 0.6}$ \\
{\bf  2DCG } & $\bf 25 $  & $\bf 195,245 $ & ${\bf 155 \pm 14}$ & \ldots & ${\bf 1.9 \pm 0.5}$ \\   
 GH  & $ 15 $ & $ 211,487 $ & $ 117\pm 20 $& $6 \pm 14$& $2.7 \pm 0.5$ \\ 
{\bf 3DS} & $\bf 15$   & $\bf 298,289 $ & ${\bf 149 \pm 18}$ & ${\bf 17 \pm 18}$ & ${\bf 1.6 \pm 0.5}$ \\
{\bf  2DCG } & $\bf 15 $  & $\bf 313,724 $ & ${\bf 148 \pm 15}$ & \ldots & ${\bf 1.5 \pm 0.5}$ \\ 
 \hline
  {\bf  WENSS} & \\ \hline
{\bf   2DCG}  & $\bf 25 $  & $\bf 92,600 $ & ${\bf 117 \pm 40}$& \ldots & ${\bf 2.9 \pm 1.9}$ \\
\hline
expected\\
\hline 
NVSS  &\ldots &\ldots & $168$ & $- 7$ & $0.48 \pm 0.04$ \\
WENSS & \ldots&\ldots & $168$ & $- 7$ & $0.42 \pm 0.03$ \\
\hline
\end{tabular}
\caption{Comparison of results. Radio dipole from NVSS: BW \citep{BW02}, 
SIN \citep{Singal11} number counts, SIF \citep{Singal11} flux weighted number counts, 
SIF* corrects SIF for slope (this work),
3DS three-dimensional estimator, mask adopted from \citet{Singal11} (this work), 
2DCG two-dimensional estimator, CG mask (this work), GH \citep{Gibelyou}; 
Radio dipole from WENSS: 2DCG two-dimensional 
estimator, CG mask (this work). The expectations for a purely kinetic radio dipole are 
given at the bottom of the table. \label{Comparison}}
\end{table}

For the amplitude of the radio dipole, the situation is more contrived. Here we focused on the study 
of linear estimators and showed that all linear estimators under investigation returned a biased estimate 
of the amplitude. The amplitude estimators of \cite{BW02} and \cite{Gibelyou} are unbiased, but the latter 
one uses a biased direction estimate as an input and is thus of limited interest. 
Besides bias, we identified another effect that reduces the dipole amplitude found by the 
flux estimator used in \citet{Singal11}.  We reduced the result of this estimator by a factor of $1.4$, due to 
the fact that the appropriate exponent of the differential number count is given by $\tilde{x} = x +1$ 
(see section \ref{xProblem}). With this correction, the result of the flux weighted estimator (denoted by SIF* in table \ref{Comparison}) is now in agreement with the 
result of \citet{BW02}.

Our  three dimensional estimate with the masking of \citet{Singal11} gives rise to
$d = ( 1.8 \pm 0.6) \times 10^{-2}$. One should keep in mind that this is a biased result, thus one cannot 
naively compare it to the expected amplitude. In order to figure out if our result is consistent with the null 
hypothesis that the radio sky is statistically isotropic, modified by the kinetic effects of our proper motion
(measured via the CMB dipole), we performed 100,000 Monte Carlo simulations. The corresponding 
histogram is shown in figure \ref{HistoBoth}. We find that only  21 of those realizations contain 
a dipole higher than the measured one  and thus we can exclude that the estimated radio dipole is just 
due to our proper motion and amplitude bias at $ 99.6 \%$ CL. This is actually very puzzling, 
as the direction of the radio dipole agrees with the direction of the CMB dipole within the 
measurement error.

We can redo this analysis with the null hypothesis that the radio dipole was accurately measured 
by \citet{BW02}. This time we find that 3402 out of 100,000 realisations are higher than our measured 
dipole. If we increase the implemented velocity towards the upper one sigma bound of the dipole 
amplitude from \citet{BW02}, we observe every  sixth simulation to be above our own measurement 
($ 16 \%$). Therefore our result is in agreement with \citet{BW02}.

Before we turn to the discussion of potential explanations, let us inspect the dipole amplitudes 
from the two dimensional estimator.  The dipole amplitudes estimated with the two dimensional estimator are 
also in agreement with the results of \citet{Singal11} and \citet{BW02}. We find $d \sin \vartheta_d = 
(1.9\pm 0.5) \times 10^{-2}$ for the NVSS analysis and $d \sin \vartheta_d = 
(2.9\pm 1.9) \times 10^{-2}$ for WENSS, which translate into lower limits on $d$. Thus, the results from the WENSS catalogue are in agreement with 
the radio dipole found in the NVSS catalogue. This is encouraging as they are prepared at different 
instruments and probe different frequencies.

\begin{figure}[!h]
\begin{center}
 \includegraphics[width=6cm,angle=270]{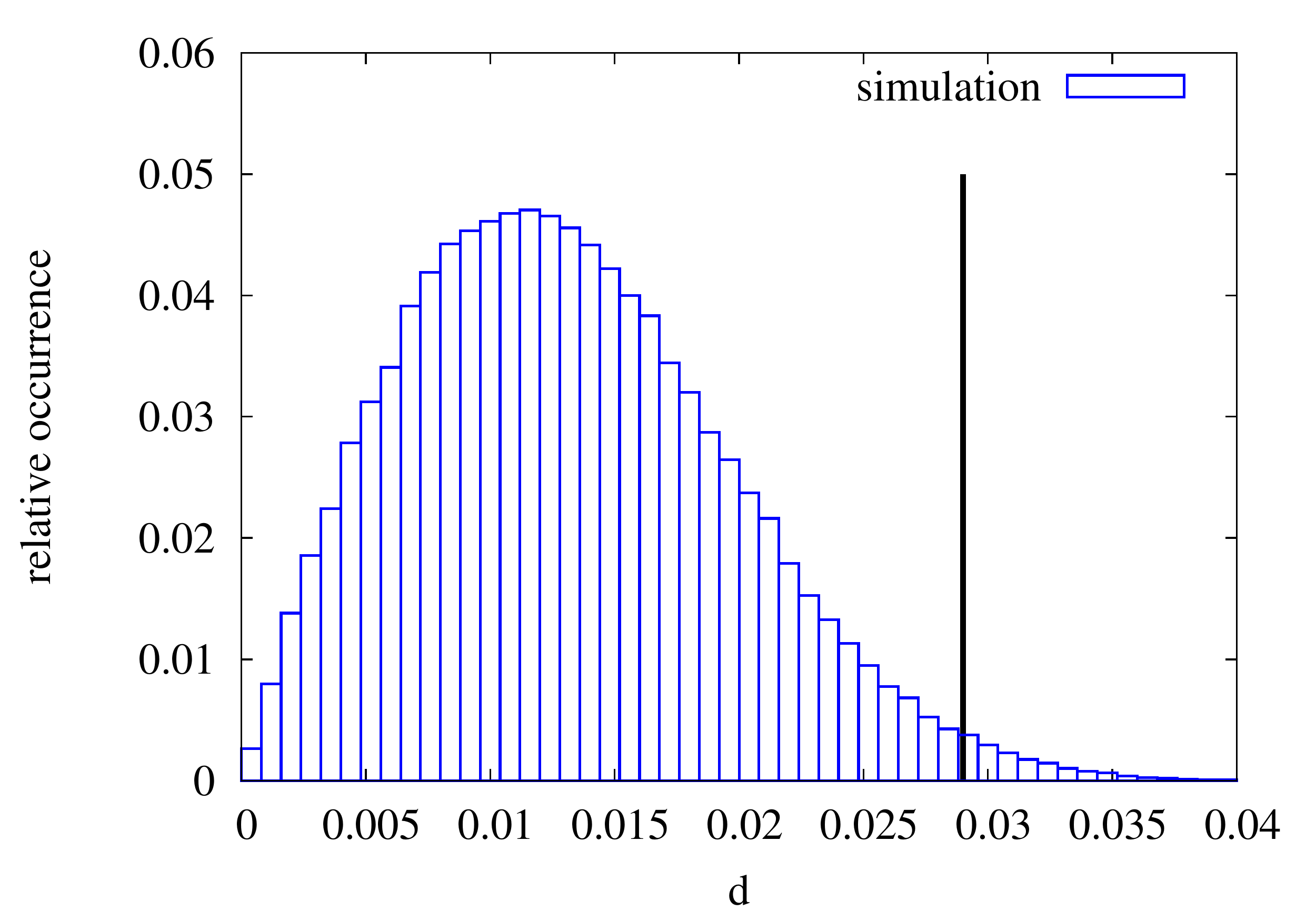}
\caption{Histogram of dipole amplitudes from 100,000 simulations for the dimensional estimator, 
assuming the CMB expectation, a slope of $x=0.8$, 92,482 sources per simulation and the CG masking 
form for the WENSS catalogue. The black vertical line is the WENSS result of this work.}\label{HistoWENSS}
\end{center}
\end{figure}

In figures \ref{HistoBoth} and \ref{HistoWENSS} we present the corresponding results from 
100,000 Monte Carlo simulations for the geometries of the NVSS and WENSS two dimensional 
estimators. In both cases we find that the amplitude bias is not enough to explain the difference between 
the observed and the expected amplitude. In the case of NVSS the null hypothesis (isotropic sky plus 
proper motion dipole) is ruled out at $99.6 \%$ CL, while for the WENSS analysis it is inconsistent at 
$98.1 \%$ CL.

The two dimensional estimation (2DCG), using the NVSS catalogue, is also compared to 
simulations, assuming the results from \citet{BW02}. We observe 1141 and 7024 out of 100,000 
simulations to give a dipole above our measurement for the assumptions of 
$d_{\rm{true}}=1.1 \times 10^{-2}$ and $d_{\rm{true}}=1.4 \times 10^{-2}$, respectively. Therefore, 
the result of our two dimensional estimator is not in contradiction to \citet{BW02}.

\section{Conclusion}
\label{Conclusion}

We conclude that the task to measure the cosmic radio dipole remains interesting and we expect that 
interesting information on cosmology can be extracted from this measurement. All measurements 
so far point towards a higher radio dipole amplitude than expected, when we assume that the cosmic 
radio dipole is just due to our peculiar motion with respect to the rest frame defined by the CMB. 
This is quite puzzling, as the orientation of the radio dipole agrees with the orientation of the CMB dipole
within measurement errors. This is the case for the NVSS and the WENSS analysis, two 
radio point source catalogues that cover $\sim 3\pi$ and $\sim \pi$ of the sky, respectively. They contain 
information at different frequencies (1.4 GHz and 325 MHz) and have been put together by different 
instruments and thus provide a strong constraint on systematic issues.  

Our detailed analysis of various linear dipole estimators \citep{Crawford, Singal11, Gibelyou} for 
three dimensional estimates $(\alpha_d, \delta_d, d)$ and a new linear estimator for a two 
dimensional estimate $(\alpha_d, d \cos \delta_d)$ had to  tackle several non-trivial issues. 
We investigated issues of directional bias, amplitude bias and
masking. There is still room to optimize the masking of the galaxy. We did not look into quadratic 
estimators, as used by \citet{BW02}. Our studies did not incorporate the uncertainties in point 
source positions, as we expect that they are subdominant (their magnitude is well below the
effect of aberration, see e.g.~\citet{WENSS}).
The measurement error of the flux is also expected to be subdominant as we 
included sources with flux above $15$ mJy only. In the case of NVSS these are a factor 6 above 
the $5\sigma$ point source detection limit, in case of WENSS it is a factor of $1.4$. The number
of point sources considered in our analysis is about $190,000$ for NVSS and $92,000$ for WENSS.
Putting all those facts together, we consider the NVSS analysis to be more reliable. Nevertheless, the 
results of the WENSS analysis are fully consistent with our results from NVSS. 

Our result (3DS) for the NVSS catalogue is $(\mathrm{RA}, \mathrm{dec}, d)=(154^\circ \pm  19^\circ,
 -2^\circ \pm  19^\circ, ( 1.8 \pm 0.6) \times 10^{-2})$.
Thus we conclude that the observed amplitude of the radio dipole exceeds the expected amplitude 
by about a factor of four. We could imagine that this might be due to the structure that causes our proper 
motion, which in a simple model of our Hubble patch would certainly be aligned with the direction of 
proper motion. However, all attempts to identify a ``great attractor'' by means of other observations
(optical, infra-red, X-ray) have failed so far. Of all those probes, the radio surveys are definitely the 
deepest probe of the Universe, as the mean redshift of NVSS sources is estimated to be 
$1.2$ by \citet{Zotti10} and $1.5$ by \citet{Ho08}. In order to explain the observed excess
radio dipole by contributions from local structure, we would need a density contrast of order 
$0.05$ at scales extending to about $z \approx 0.3$, which does not seem plausible.
Without a detailed study of the redshift distribution of the radio sources it is impossible to judge 
whether this finding is actually in agreement with the current standard model of cosmology.  

An example of such a scenario is provided by \citet{DavidW}.
They claim that the spherically averaged Hubble law on $ < 100/h \  Mpc$ scales
is significantly closer to uniform in the Local Group frame as compared to
the CMB frame and on this basis have suggested a non-kinematic contribution to
the CMB dipole. In this case the CMB dipole could differ from
the cosmic radio dipole. 

Another reason for the large amplitude of the radio dipole could be that the linear estimators considered 
in this work do not assume the deviation from isotropy to be a pure dipole. Thus higher multipole
moments also contribute to the measured amplitude. 

It has been found from the analysis of the CMB 
that quadrupole, octopole and a few more low $\ell$-multipoles seem not to be orientated randomly on 
the sky, but show some unexpected alignments \citep{Schwarz, Bennett, Copi} among themselves and 
with the CMB dipole direction. Thus it might not be surprising that also the dominant anisotropy direction of 
the radio sky lines up.  
    
It is evident that it would be crucial to repeat this investigation with new and even deeper 
radio catalogues, which provide more sources. In the near future there will be three large sky 
surveys available \citep{Raccanelli}. A multi-wavelength study will be possible based on the 
Multifrequency Snapshot Sky Survery (MSSS) of the International LOFAR Telescope and with the 
LOFAR Tier-1 survey. The Australian Square Kilometre Array (SKA) Pathfinder (ASKAP) will 
produce the Evolutionary Map of the Universe (EMU) and the Westerbrok Synthesis Radio 
Telescope (WSRT) equipped with the Apertif system will compile the Westerbrok Observations of 
the Deep Apertif Northern sky survey (WODAN) catalogue. 

The multi-wavelength surveys will also allow us to directly measure the spectral index $\alpha$, which has to 
be known to connect the measured amplitude to the kinetic dipole. A steepening of the spectral index 
for the lowest flux sources would increase the expected kinetic amplitude.  HI surveys will have the 
advantage that they will also provide redshift information on top of positions and fluxes and we will be 
able to study the evolution of the radio dipole as a function of redshift out to redshifts of a few. 
Beyond that SKA will increase the number of sources in such a survey by orders of magnitude. 
All these surveys will reduce the random dipole contribution, improve on systematics, and allow us to 
settle the question: Is the radio dipole in agreement with the CMB dipole?

\begin{acknowledgements} 
We thank David J. Bacon, Dragan Huterer, Matt J. Jarvis, Huub R\"ottgering, Seshadri Nadathur and Ashok K. Singal for fruitful 
comments and discussions and Anne-Sophie Balleier for help with Healpy and the generation of maps 
of the NVSS and WENSS number counts. The CAMB code, and the Healpy and HEALPix 
\citep{Gorski} packages 
have been used to estimate the concordance value of the CMB primordial dipole and to produce the 
NVSS and WENSS maps of the sky. We acknowledge financial support from Deutsche 
Forschungsgemeinschaft (DFG) under grants IRTG 881 `Quantum Fields and Strongly Interacting 
Matter' and RTG 1620 `Models of Gravity', as well as from the Friedrich Ebert Stiftung.
\end{acknowledgements}

\appendix
\section{Monte Carlo Simulations}

We used the pseudorandom number generator Mersenne Twister for all Monte Carlo simulations. One simulated source consists of two position coordinates and one flux value. The coordinates will be drawn from an uniform distribution, leading to an isotropic sky for a catalogue with many sources. In order to obtain a desired number count power law like (\ref{numbercountpowerlaw}) with a certain slope $x$, we calculate the flux $S$ using a random number $A$ (choosen between 0 and 1) by
\begin{equation}
   S_{\rm{rest}}= S_L (1-A)^{-x}  , 
\end{equation}
 where $S_L$ is a flux value $20 \%$ below the simulated flux limit. The simulation also creates sources, which only due to Doppler shifting are counted in the final catalogue, because this value of $S_L$ lies below the simulated flux limit. Lowering $S_L$ further increases computational time and is not necessary, as long as the simulated velocity $v$ is below $0.1 c$.
 
The two physical effects (Doppler shift and spherical aberration) will be implemented separately.
In cooperating the Doppler effect is straightforward, since it only affects the flux values of each source depending on the angle $\theta$ between the source and the velocity direction, i.e.
\begin{equation}
S_{\rm{obs}}(\nu_{\rm{obs}})=S_{\rm{rest}} \left( \frac{1+\frac{v}{c} \cos(\theta)}{\sqrt{1-(\frac{v}{c})^2}} \right)^{1+\alpha} .
\end{equation}
In the simulation $\alpha$ is fixed to $0.75$ for all sources. The velocity direction and amplitude can by chosen be the user.

To model the relativistic effect of stellar aberration one has to change the position of each radio source. The aberration formula is 
\begin{equation}
 \tan(\theta^\prime)=\frac{\sin(\theta)\sqrt{1-(\frac{v}{c})^2}}{\frac{v}{c}+\cos(\theta)} ,
\end{equation}
where $\theta^\prime$ is the new angle between the velocity direction and the radio source. 
Forth, the position of a radio source is translated into Cartesian coordinates by assuming that it lies on a unit sphere. Then a straight line from this point ($\vec P$) to the velocity direction on the sphere ($\vec V$) is constructed depending on a parameter $t$
\begin{equation}
 \vec r(t)= \vec P (1-t)+ \vec V t .
\end{equation}
On this line we choose a $t^\prime$ in such a way that $\vec{r}(t^\prime)$ points towards the new position. The value of $t^\prime$ can be determined by
\begin{eqnarray}
 \vec r(t^\prime) \cdot \vec V = r^\prime(t^\prime) \cos(\theta^\prime)  \\ \rightarrow t^\prime = \frac{r^\prime(t^\prime) \cos(\theta^\prime)-\cos(\theta)}{1-\cos(\theta)}  
\end{eqnarray}
with $r^\prime(t^\prime)=\sqrt{\vec{r}^{\prime 2}(t^\prime)}$. This equation is solved by
\begin{equation}
  t^\prime_1 = \frac{\sin(\theta-\theta^\prime)}{\sin(\theta - \theta^\prime)+\sin(\theta^\prime)} \vee t^\prime_2 = \frac{\sin(\theta + \theta^\prime)}{\sin(\theta + \theta^\prime)-\sin(\theta^\prime)}  .
\end{equation}
 We know that for $\theta=\theta^\prime$ the result of $t^\prime$ must always be $0$. Therefore the correct solution is $t^\prime=t^\prime_1$. Now one has to transform $\vec r(t^\prime)$ back into spherical coordinates in order to find the new position of the radio source. The new declination $\vartheta^\prime$ is then (the index v stands for the velocity direction):
\begin{equation}
 \vartheta^\prime = \arccos \left(\frac{1}{r^\prime}[(1-t^\prime)\cos(\vartheta)+t^\prime \cos(\vartheta_{\rm{v}})]\right)
\end{equation}
and the new right ascension $\varphi^\prime$:
\begin{equation}
 \varphi^\prime=\arcsin \left( \frac{(1-t^\prime) \sin(\vartheta)\sin(\varphi)  + t^\prime \sin(\vartheta_{\rm{v}}) \sin(\alpha_{\rm{v}})} {r^\prime \sin(\vartheta^\prime)} \right)  .
\end{equation}
This way one obtains a simulated sky, including the effect of the observers movement. Now one can feed the different estimators with those sky simulations and obtain the resulting dipole vectors.

\bibliographystyle{aa} 
\bibliography{dipole}

\end{document}